\documentclass[twocolumn]{aastex631}
\hypersetup{linkcolor=blue,citecolor=blue,filecolor=cyan,urlcolor=blue}

\newcommand{\IZw}{\rm I\,Zw\,18 }

\shorttitle{The ULX in the metal-poor galaxy I\,Zw\,18}

\graphicspath{{./}{figures/}}
\begin{document}
\title{Possible Supercritical Accretion on the ULX in the Metal-poor Galaxy I\,Zw\,18}
\correspondingauthor{Marina Yoshimoto}
\email{myoshimo@ess.sci.osaka-u.ac.jp}
\author[0009-0005-0819-0819]{Marina Yoshimoto}
\affiliation{Department of Earth and Space Science, Graduate School of Science, Osaka University, 1-1 Machikaneyama, Toyonaka, Osaka 560-0043, Japan}
\author[0000-0002-2683-6856]{Tomokage Yoneyama}
\affiliation{Department of Physics, Faculty of Science and Engineering, Chuo University,1-13-27 Kasuga, Bunkyo-ku, Tokyo 112-8551, Japan}
\author[0000-0001-6020-517X]{Hirofumi Noda}
\affiliation{Astronomical Institute, Tohoku University, 6-3 Aramakiazaaoba, Aoba-ku, Sendai, Miyagi 980-8578, Japan}
\author[0000-0003-2670-6936]{Hirokazu Odaka}
\affiliation{Department of Earth and Space Science, Graduate School of Science, Osaka University, 1-1 Machikaneyama, Toyonaka, Osaka 560-0043, Japan}
\affiliation{Forefront Research Center, Graduate School of Science, Osaka University, 1-1 Machikaneyama, Toyonaka, Osaka 560-0043, Japan}
\affiliation{Kavli Institute for the Physics and Mathematics of the Universe, The University of Tokyo, 5-1-5 Kashiwanoha, Kashiwa, Chiba, 277-8583, Japan}
\author{Hironori Matsumoto}
\affiliation{Department of Earth and Space Science, Graduate School of Science, Osaka University, 1-1 Machikaneyama, Toyonaka, Osaka 560-0043, Japan}
\affiliation{Forefront Research Center, Graduate School of Science, Osaka University, 1-1 Machikaneyama, Toyonaka, Osaka 560-0043, Japan}
\begin{abstract}
We present an analysis of X-ray observations of the Ultraluminous X-ray source (ULX) in \IZw based on archival data taken with {\it Chandra}, {\it XMM-Newton}, and {\it Suzaku}. 
This ULX is considered to be an intermediate-mass black hole candidate simply because it is in the lowest metallicity environment among ULXs, where formation of heavy black holes is facilitated. 
However, actual study of the ULX based on observations spanning for a long period has been too limited to determine its nature. 
In this study, we investigate the spectral evolution of the ULX up to 2014, combining the previously-unpublished {\it Suzaku} data with those from the other two satellites.
We derive a positive correlation of $L\propto T_{\rm in}^{2.1\pm0.4}$ between the bolometric luminosity $L$ and inner-disk temperature $T_{\rm in}$ on the basis of the multi-color disk-blackbody model, where we exclude the {\it Chandra} data, which has the lowest luminosity and systematic residuals in the fitting.  The nominal relation $L\propto T_{\rm in}^{4}$ for the standard disk is rejected at a significance level of 1.5\,\%.
These results suggest that the ULX was in the slim-disk state during these observations except at the time of the {\it Chandra}  observation, in which the ULX was likely to be in a different state. 
The apparent inner-disk radius appears negatively correlated with the inner-disk temperature.
Moreover, we find a radial dependence of the disk temperature of $T (r)\propto r^{-p}$ with $p<0.75$, which also supports the hypothesis that the ULX has a slim disk.
In consequence, the \IZw ULX is most likely to be powered by a stellar-mass compact object in supercritical accretion.
\end{abstract}
%
\section{Introduction} \label{sec:intro}
Ultraluminous X-ray sources (ULXs) are generally known to be non-nuclear, point-like objects observed in external galaxies. In particular, the most distinctive feature is a soft X-ray $({\rm 0.3-10\,keV})$ luminosity of $L_{\rm X}\gtrsim10^{39}\,{\rm erg\,s^{-1}}$, which exceeds the Eddington limit for stellar-mass black holes (BHs) of $10\,M_{\rm \odot}$, under an assumption of isotropic emission \citep[see reviews by][]{2011NewAR..55..166F, 2017ARA&A..55..303K,2023NewAR..9601672K}. 
The discovery of ULXs was made with the {\it Einstein} Observatory \citep{1989ARA&A..27...87F}, and 1843 candidates, according to \citet{2022MNRAS.509.1587W}, are known by now.  
The soft X-ray spectra of ULXs are typically reproduced with a model of multi-color disk blackbody (MCD) and Comptonization components, as observed in Galactic black hole X-ray binaries (GBHBs). 
However, the presence of a spectral cutoff above $10\,{\rm keV}$ in ULX spectra \citep{2013ApJ...778..163B} is a distinctively different feature from GBHBs.
The luminosity of an accretion disk strongly depends on the mass and mass accretion rate of the object.  Accordingly, the ultra-high\ luminosity of ULXs, much higher than that of  GBHBs, is usually hypothesized with either or both massive central objects and/or supercritical accretion.\par
The earliest theories proposed that ULXs were a population of intermediate-mass black holes (IMBHs) with sub-Eddington accretion \citep{1999ApJ...519...89C, 2000ApJ...535..632M}. 
For instance, ESO 243-49 HLX-1,  one of the so-called hyper-luminous X-ray sources, which are the ULX population defined with a luminosity of $L_{\rm X}\gtrsim10^{41}\,{\rm erg\,s^{-1}}$  \citep{2003ASPC..289..291M}, has been an IMBH candidate on the basis of the relation between luminosity and inner-disk temperature consistent with a standard disk property of $L_{\rm bol}\propto T^{4}_{\rm in}$, where $L_{\rm bol}$ is a bolometric luminosity \citep{1973A&A....24..337S}, at a luminosity of $\sim10^{42}\,{\rm erg\,s^{-1}}$.
However, the discovery of pulsation \citep{2014Natur.514..202B} and ultrafast ($\sim0.2\,c$) outflows \citep{2016Natur.533...64P,2017MNRAS.468.2865P} in some ULXs suggests that stellar-mass BHs or neutron stars (NSs) with accretion in the super-Eddington mode are also candidate ULXs. 
Therefore, sub-Eddington accreting source is not the only nature of ULXs.
ULXs are simply defined by the observed luminosity, and \citet{2016AN....337..349B}  suggested that a population of ULXs  is in fact a ``zoo'' with a variety of species.\par
The majority of ULXs are found in spiral and irregular galaxies with high star-formation rates \citep[SFRs; e.g., ][]{2004ApJ...601L.143I}. 
The number of ULXs in a galaxy normalized by the SFR is anti-correlated with metallicity \citep{2002astro.ph..2488P,2010MNRAS.408..234M,2011AN....332..414M,2013ApJ...769...92P}.
In the early universe, which had zero metallicity, deficiency of heavy elements is thought to facilitate forming of massive stars \citep[Population III; e.g.,][]{2014ApJ...781...60H}.
The massive stars eventually become heavy BHs \citep[e.g.,][]{2010ApJ...714.1217B}.
The environment also facilitates accretion of Roche-lobe overflow type onto compact objects in binary systems \citep{2010ApJ...725.1984L}.
For these reasons, it may be possible that ULXs evolved in low metallicity galaxies are driven by IMBHs in sub-Eddington accretion \citep{2009MNRAS.400..677Z,2010MNRAS.408..234M}. 
\citet{2017ApJ...846...17W} proposed that the nature of ULXs is expected to differ depending on the star formation history of their host galaxies.
If the SFR  does not vary over time in low metallicity environments, BH ULXs are more abundant than NS ULXs during 10\,Gyr since the beginning of star formation in the Universe.
Conversely, if a burst of star-formation occurred in the first 100\,Myr, NS ULXs should have outnumbered BH ULXs within $\lesssim\,1\,{\rm Gyr}$ and have become dominant after a few Gyr, regardless of the metallicity.
Determining the nature of the ULX in metal-poor galaxies is crucial for delving into the evolution of binary systems.
\par
Blue Compact Dwarf galaxies (BCDs) are the least chemically evolved galaxies in the local universe \citep{1981ApJ...247..823T}, which has an extraordinarily low metallicity of $1/50-1/3 \,Z_{\odot}$ \citep[see, e.g., the review by][]{2000A&ARv..10....1K}.
\citet{2011ApJ...741...10K} revealed the total X-ray luminosity of four X-ray point sources in three BCDs is 10 times larger than that of star-forming galaxies with normal metallicity. 
The finding suggests that the low metallicity environment is closely related to the evolution of X-ray sources. \par
 I\,Zwicky\,18 \citep[I\,Zw\,18;][]{1966ApJ...143..192Z} is an extreme galaxy in BCDs.
Optical spectroscopy found that the amount of oxygen was ${\rm 12+log(O/H)=7.11\pm0.01}$ as a representative value \citep{2016MNRAS.459.2992K}. Comparing this with the solar value of ${\rm 12+log(O/H)=8.69\pm0.05}$ \citep{2009ARA&A..47..481A}, the abundance of $Z=0.026\,Z_{\rm \odot}$ is derived. 
Previously, \IZw had been thought to be composed of young systems that started star-formation only 500\,Myr ago \citep{2003ApJ...588..281H,2004ApJ...616..768I}. However, \citet{2007ApJ...667L.151A} indicated on the basis of deep HST/ACS photometry that \IZw contains old stars (${\rm \geqq1\,Gyr}$), the fact of which implies that star formation started at $z\geqq0.1$. 
The distance for I\,Zw\,18 was obtained as $18.2\pm1.5\,{\rm Mpc}$ and $19.0^{+2.8}_{-2.5}\,{\rm Mpc}$  from tip of red giant branch \citep{2007ApJ...667L.151A} and Cepheids \citep{2010ApJ...711..808F,2010ApJ...713..615M}, respectively. We adopted the distance of 18.2\,Mpc in this work as well as previous works \citep{2013ApJ...770...20K,2017A&A...602A..45L,2021ApJ...908L..54K}.
\par
\IZw contains an ULX located at ${\rm R.A.=09^{h}34^{m}02\fs0}$, ${\rm DEC.=+55\arcdeg14\arcmin28\farcs0\,(J2000.0)}$.
In the past decades, X-ray spectral modeling of the ULX was carried out in several previous works.
\citet{1996rftu.proc..375F} showed the X-ray source has a luminosity of $6.2^{+2.3}_{-4.9}\times10^{39}\,{\rm erg\,s^{-1}}$ in the energy range of $0.1-2.4\,{\rm keV}$, using {\it ROSAT}/PSPC data.
\citet{1996ApJ...465..680M} also used {\it ROSAT}/PSPC data and reported a luminosity of $\thickapprox1\times10^{39}\,{\rm erg\,s^{-1}}$ to within a factor of two accounting for the Galactic absorption.
\citet{2002ASPC..262..141B} confirmed that the {\it Chandra}/ACIS-S data in 2000 was well fitted with a power-law model. \citet{2004ApJ...606..213T} reported a luminosity of $1.6\times10^{39}\,{\rm erg\,s^{-1}}$ from the power-law model fitting with the {\it Chandra}/ACIS-S data in the energy range of $0.5-10\,{\rm keV}$.
The first application of the MCD model, beyond the simple power-law model, to the ULX was made by \citet{2013ApJ...770...20K} with the {\it XMM-Newton}/EPIC data in 2002. 
They found a luminosity exceeding $10^{40}\,{\rm erg\,s^{-1}}$ in the energy range of $0.3-10\,{\rm keV}$ and interpreted the change in luminosity from the \textit{Chandra} observation as a hard-to-soft transition between the two observations. 
Comparing the maximum luminosities of GBHBs in the hard state with the luminosity in the {\it Chandra}  observation, the mass of the ULX was estimated to be $>85\,M_{\odot}$, suggesting that it is an IMBH.
\citet{2021ApJ...908L..54K} applied a power-law model to the {\it ROSAT}/PSPC, {\it Chandra}/ACIS-S, {\it XMM-Newton}/EPIC, and {\it Swift}/XRT data  taken from 1992 to 2012.
They reported that the luminosity of the X-ray source varied on timescales from days to years.\par
As stated above, the ULX, in the metal-poor galaxy {\rm I\,Zw\,18}, was suggested to be an IMBH candidate.
However, there was no study on the nature based on observations spanning for a long period.
In this study, we investigate the long-term spectral evolution of the ULX in \IZw up to 2014, analyzing the previously-unpublished {\it Suzaku}/XIS data taken in two occasions in 2014 and combining them to the previously published data.
{\it Suzaku}/XIS observed the \IZw ULX twice in 2014 with exposures of 17.3\,ks and 82.1\,ks.
The second one of which is the longest single observation of this ULX ever made in X-rays.
The {\it ROSAT}/PSPC and {\it Swift}/XRT data are excluded from the analysis due to insufficient photon statistics. 
We focus on disk radiation and examine variations in luminosity, inner-disk temperature, apparent inner-disk radius, and the radial dependence of disk temperature to investigate the nature of the ULX.

\section{Observations and Data Reduction} \label{sec:obsred}
We use five archival data sets made with {\it Chandra}/ACIS-S, {\it XMM-Newton}/EPIC, and {\it Suzaku}/XIS. The details of the observations are listed in Table \ref{tab:obs}. \par
\begin{figure}[th]
\epsscale{1} 
\plotone{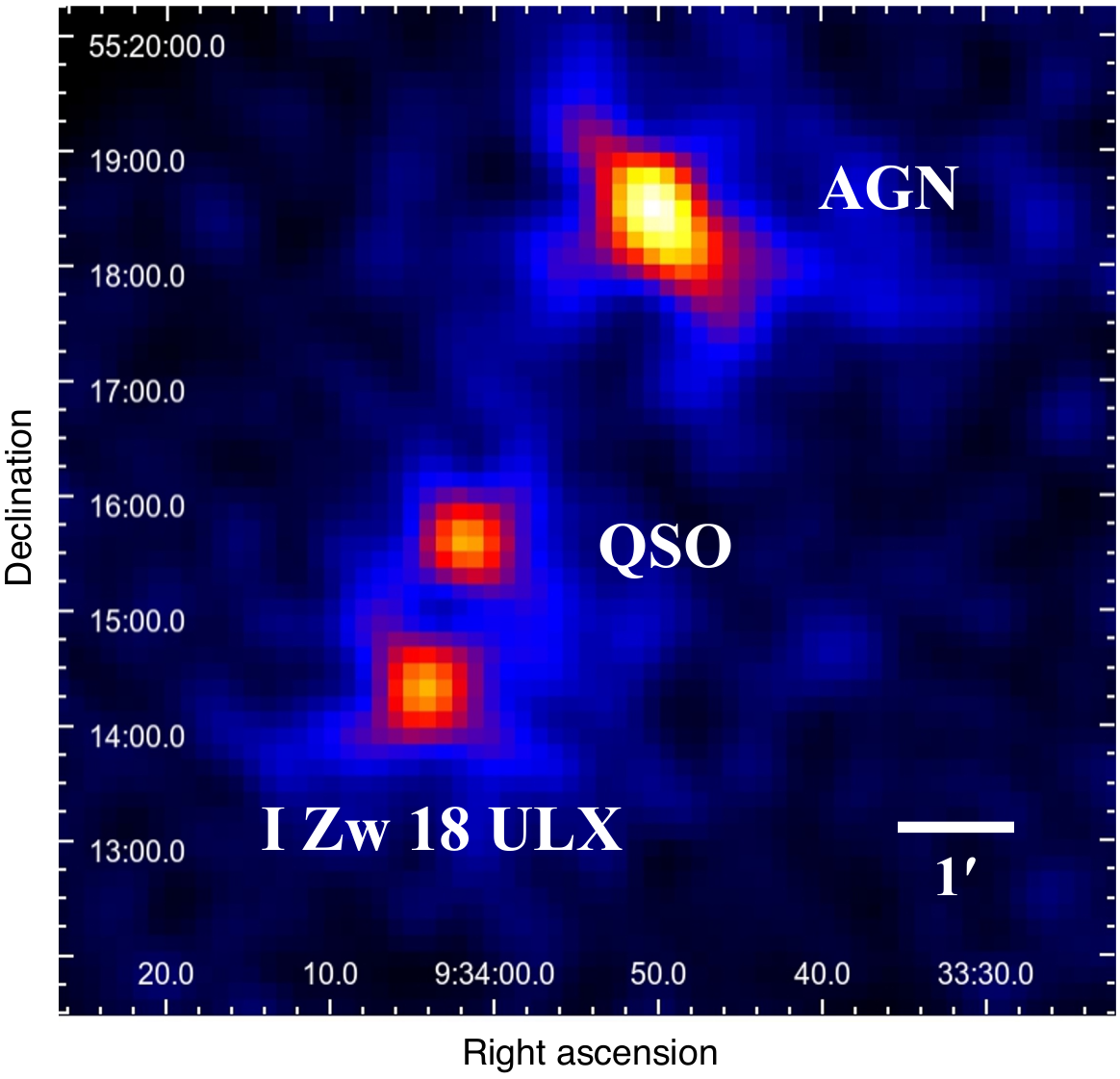}
\caption{{\it Suzaku}/XIS0 image at epoch S2 around the \IZw ULX in the energy range of $0.5-7\,{\rm keV}$. 
On the celestial plane, a QSO and AGN are located  $1\farcm4$ and $5\arcmin$ away from the ULX, respectively.
The angular distance between the ULX and QSO  is shorter than the half-power diameter of {\it Suzaku} ($\sim2\arcmin$), leading to contamination of photons between the two sources.
\label{fig:suzaku_image}
}
\end{figure}

The \IZw ULX is situated adjacent to a quasi-stellar object \citep[QSO; 2CXO J093359.3+551550 in {\it Chandra} source Catalog version 2.0][]{2020AAS...23515405E} and an active galactic nucleus \citep[AGN; \lbrack AGG2006\rbrack\, 59 in {\it Sloan Digital Sky Survey}\/{\it XMM-Newton} AGN Spectral Properties Catalog,][]{2006A&A...459..693A} on the celestial plane (Figure \ref{fig:suzaku_image}). 
They are potential contamination sources. We thus excluded them in the analysis of the {\it Suzaku} data.

\begin{deluxetable*}{ccccccc}
\tablenum{1}
\tablecaption{Observation details of the \IZw ULX analyzed in this work\label{tab:obs}}
\tablewidth{0pt}
\tablehead{
\colhead{Epoch} & \colhead{Observatory} & \colhead{Date} & \colhead{ObsID/Sequence number} & \colhead{Instrument} &\colhead{Net exposure time} &\colhead{Count rate}\\
\colhead{} & \colhead{} & \colhead{[YYYY-MM-DD]} & \colhead{} & \colhead{} &\colhead{[ks]} &\colhead{[counts ${\rm ks}^{-1}$]}
}
\startdata
C1 & {\it Chandra} & 2000-02-08 & 805/600108 & ACIS-S & 40.8 & $13.0\pm0.6$ \\
X1 & {\it XMM-Newton} & 2002-04-10 & 0112520101 & EPIC-pn & 24.1 & $91.4\pm2.1$ \\
&  &  &  & EPIC-MOS1 & 31.1 & $28.7\pm1.0$ \\
&  &  &  & EPIC-MOS2 & 30.9 & $28.7\pm1.0$ \\
X2 & {\it XMM-Newton} & 2002-04-16 & 0112520201 & EPIC-pn &  5.4 & $79.1\pm4.1$ \\
&  &  &  & EPIC-MOS1 & 18.5 & $26.7\pm1.8$ \\
&  &  &  & EPIC-MOS2 & 18.6 & $28.0\pm1.8$ \\
S1 & {\it Suzaku} & 2014-05-15 & 709021010 & XIS0\,+\,XIS3 & 17.3\,+\,17.3 & $4.2\pm0.4$ \\
&  &  &  & XIS1 & 17.3 & $4.4\pm0.8$ \\
S2 & {\it Suzaku} & 2014-10-04 & 709021020 & XIS0\,+\,XIS3 & 82.1\,+\,82.1 & $2.5\pm0.1$ \\
&  &  &  & XIS1 & 82.1 & $2.4\pm0.3$ \\
\enddata
\tablecomments{
Table columns from left to right: the name of epoch, observatory, observation start date, observation ID or sequence number, instrument, net exposure time, and count rate.
In calculating the net exposure time,  the periods with prominent flaring particle backgrounds are removed. 
The net exposure times of the EPIC-MOS data at epoch X2 are the sum of scheduled and unscheduled times.
}
\end{deluxetable*}
%
\subsection{\it Chandra}  \label{subsec:chandra}
\IZw was observed with {\it Chandra} Advanced CCD Imaging Spectrometer (ACIS) -S with an exposure of 40.8\,ks in the FAINT mode on 2000 February 8 to 9 (ObsID: 805, Sequence number: 600108, PI: Bomans, Epoch: C1). Data reduction was carried out using the {\it Chandra} Interactive Analysis of Observation software package \citep[CIAO;][]{2006SPIE.6270E..1VF} version 4.14.0 with the {\it Chandra} X-ray Calibration Database (CALDB) files version 4.9.8. The downloaded data was reprocessed with the task {\tt chandra\_repro}. 
We extracted the image with the task {\tt dmcopy} and assigned source and background regions as a circular region with a radius of $5\arcsec$ and an annulus with radii of ${\rm 5\arcsec-30\arcsec}$, respectively. 
The spectra, a photon redistribution matrix file (RMF), and an ancillary response file (ARF) were created using the task {\tt specextract}. 
Our study identified background flaring in light curves generated with the task {\tt dmextract}, as well as \citet{2004ApJ...606..213T}. 
However, the time regions that encompass background flaring are included in the following analysis because we found it does not significantly affect our results. The energy range for our spectral fitting is ${\rm0.4-10\,keV}$. 
 %
\subsection {\it XMM-Newton} \label{subsec:xmm-newton}
We analyzed the archival data of two {\it XMM-Newton} observations in the Imaging mode on 2002 April 10 to 11 (ObsID: 0112520101; PI: Watson, Epoch: X1), and 2002 April 16 to 17 (ObsID: 0112520201; PI: Watson, Epoch: X2). 
The observation data of the pn, MOS1, and MOS2 detectors of the European Photon Imaging Camera (EPIC)  were processed with the Science Analysis System (SAS) version 20.0.0 in the standard procedure. 
The calibration files of the March 2022 version were applied to the data with the tasks {\tt epchain} and {\tt emchain} for pn and MOS, respectively.
We selected single and double events ({\tt PATTERN<=4}) only for the pn and single, double, triple, and quadruple events ({\tt PATTERN<=12}) only for the MOSs, with flags of {\tt \#XMMEA\_EP \&\& FLAG==0} and {\tt \#XMMEA\_EM}, respectively.
The periods with prominent flaring particle-backgrounds were determined from the light curves generated with the task {\tt evselect} for single events in the high energy range ($10-12\,{\rm keV}$ for pn, $>10\,{\rm keV}$ for MOSs).
With these high-background periods excluded, the resultant net exposure times for pn, MOS1, and MOS2 were, respectively, 24.1, 31.1, and 30.9\,ks in epoch X1, and 5.4, 18.5, and 18.6\,ks in epoch X2. 
After these time-filtering, we extracted the images and spectra with the task {\tt evselect}.
The source regions for our analyses were a $30\arcsec$-radius circle for each detector.
The background regions for pn were three regions on the CCD chip encompassing the source, two circular regions of a $30\arcsec$-radius (east and west) and a circular region with a radius of $70\arcsec$ (south).
Those for MOSs were, from an annulus region with radii of $30\arcsec-180\arcsec$ excluding the  $30\arcsec$-radius circle centered at the QSO. 
In the MOS1 and MOS2 data in epoch X2, the spectra for the scheduled and unscheduled observations were extracted separately.
We created RMFs and ARFs for each observation using the tasks {\tt rmfgen} and {\tt arfgen}, respectively. The spectral energy range is $0.2-10\,{\rm keV}$.
%
\subsection {\it Suzaku} \label{subsec:suzaku}
Two pointing observations with \textit{Suzaku} were made with the X-ray Imaging Spectrometers (XIS) 0, 1, and 3 on 2014 May 15 (ObsID: 709021010; PI: Kaaret, Epoch: S1) and 2014 October 4 to 5 (ObsID: 709021020; PI: Kaaret, Epoch: S2). 
For epoch S1, there were only $3\times3$ cleaned events for each detector with an exposure time of ${\rm 17.3\,ks}$. 
For epoch S2,  we added $5\times5$ events to $3\times3$ events, resulting in a total exposure time of {\rm 82.1\,ks} for each detector. 
Images were extracted using the task {\tt xselect} in HEASoft \citep{2014ascl.soft08004N} version 6.30.1. \par
In the images, we found that the positions of the ULX, QSO, and AGN were shifted in the same direction from the catalog value. 
The typical positional uncertainty of {\it Suzaku} is $19\arcsec$ \citep{2008PASJ...60S..35U}. 
We corrected the celestial positions of objects based on the cataloged position of the AGN. 
The source spectra were accumulated from a circular region of 41\arcsec-radius, using the task {\tt xselect}. 
The background spectra were accumulated from an annular region of 2\arcmin\,to 6\arcmin\, radii centered at the middle position between the ULX and QSO, where
 a circular region of 1\farcm8-radius centered at the AGN was excluded.
\par
The RMFs were created with the task {\tt xisrmfgen} and the ARFs were, with the task {\tt xissimarfgen} \citep{2007PASJ...59S.113I} with CALDB files of the October 2018 version. 
The ULX and QSO are only $1\farcm4$ apart on the celestial plane. Given that the X-ray telescope (XRT) has a half-power diameter (HPD) of $\sim2\arcmin$ \citep{2007PASJ...59S...9S}, 
 the contamination from the QSO source due to the point spread function is significant and hence should be taken into account when analyzing the ULX data. 
The fraction of the photons leaking from the ULX and QSO into each other was reproduced through creating ARFs for ULX and QSO.
For the front illuminated CCDs, XIS0 and XIS3, the spectra were summed together using the task {\tt mathpha}, and response files were compiled by weighting the exposure time.
We made simultaneous model fitting with the spectra of ULX and QSO in the energy range of $0.5-10\,{\rm keV}$. 
%
\section{Analysis and Results} \label{sec:results}
The spectra are grouped into a minimum of 20 counts per energy bin using the HEASoft task {\tt grppha}, the CIAO task {\tt specextract}, and the SAS task {\tt specgroup} for {\it Suzaku}, {\it Chandra}, and {\it XMM}, respectively. Simultaneous spectral model-fitting is performed with ${\chi^2}$ statistics for the spectra of all the detectors in each epoch, in which the background is subtracted, using XSPEC version 12.12.1 \citep{1996ASPC..101...17A}.  
In this section, the quoted errors of fitting parameters are  in a 90\,\% confidence level unless mentioned otherwise. \par
\begin{figure*}[th]
\epsscale{1} 
\plotone{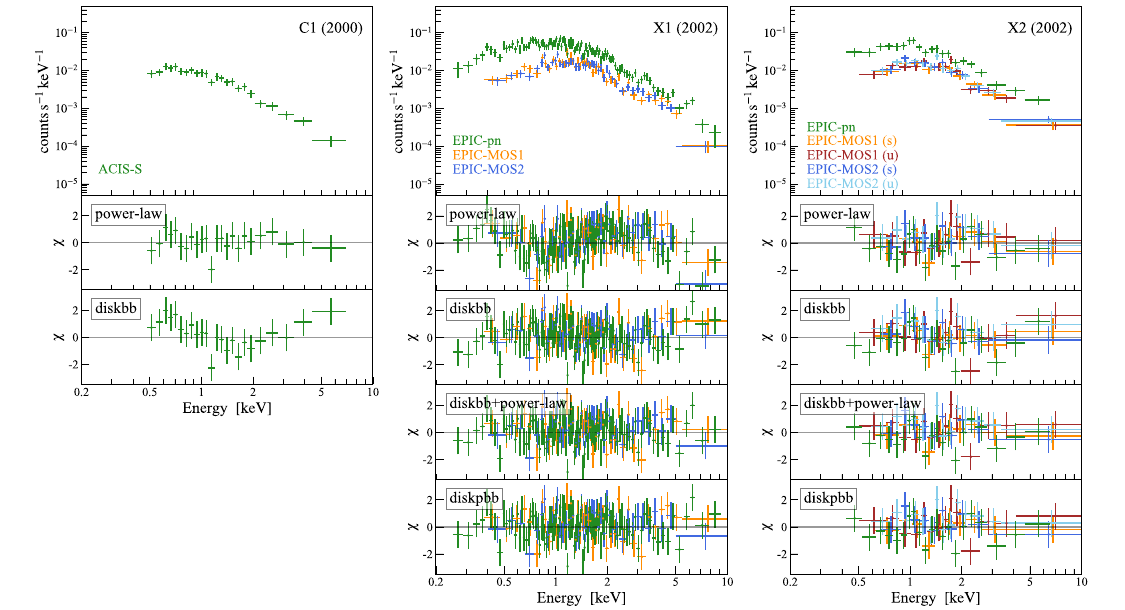}
\caption{The $0.2-10\,{\rm keV}$ spectra of epochs C1, X1, and X2 for the \IZw ULX. 
Each data point is plotted with an error in 1\,$\sigma$ confidence level.
The $\chi\,{\rm (=data-model/error)}$ distributions for absorbed models of {\tt powerlaw}, {\tt diskbb}, {\tt diskbb+powerlaw}, and {\tt diskpbb} (if applied) are shown in the lower panels below the spectra. 
For the EPIC-MOS1 and MOS2 data in epoch X2, the data in the scheduled (labeled as (s)) and unscheduled (u) periods are treated as the separate spectral data sets and are fitted simultaneously in combination with the pn data.
\label{fig:spectra_prev}}
\end{figure*}
\begin{figure*}[th]
\epsscale{1} 
\plotone{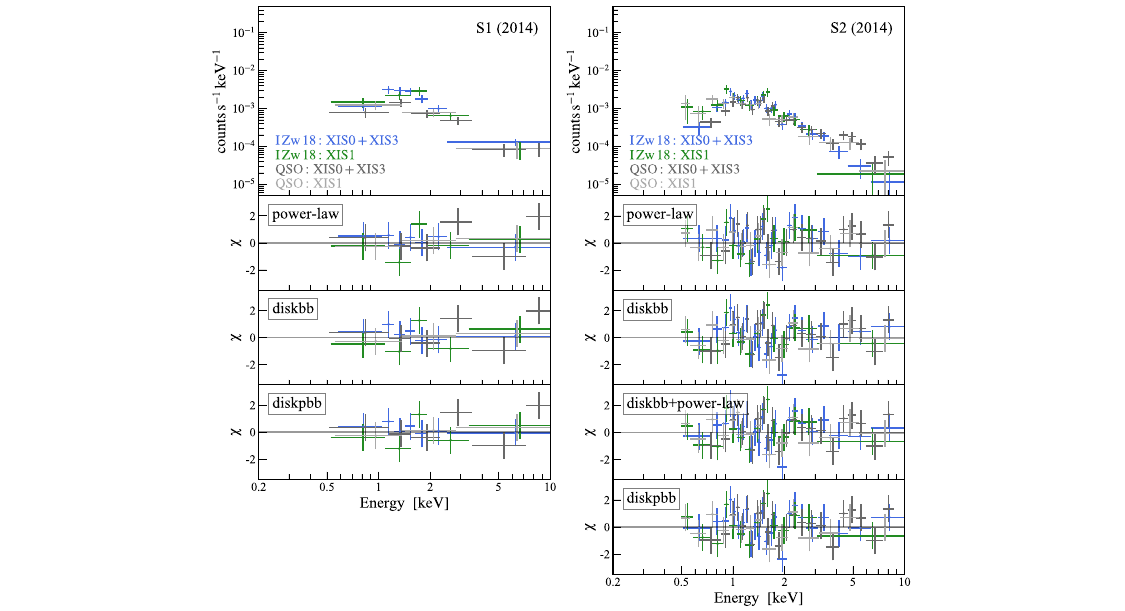}
\caption{ The same as in Figure~\ref{fig:spectra_prev} but with the {\it Suzaku}/XIS data in epochs S1 and S2, the data of which have never been published before. 
The \IZw ULX and QSO are $1\farcm4$ apart on the celestial plane, the distance of which is smaller than the {\it Suzaku} HPD.
\label{fig:spectra_suzaku}}
\end{figure*}

We estimate the Galactic absorption value to be $N_{\rm H}=2.92\times10^{20}\,{\rm cm^{-2}}$, using  the HEASoft task {\tt nh} \citep{2016A&A...594A.116H} and convolve it as the {\tt TBabs} model in XSPEC.
The absorption intrinsic to the ULX is represented with the {\tt TBvarabs} model and is free to vary with fixed parameter values of an abundance of $Z=0.026\,Z_{\odot}$ and a redshift of $z=0.00254$. 
The spectral fitting results are listed in Table 2. All the data spectra, best-fit models,  and residuals between them are shown in Figures \ref{fig:spectra_prev} and \ref{fig:spectra_suzaku}. 
In the case of the {\it Suzaku} data, the \IZw ULX and QSO spectra are fitted simultaneously, and then the QSO spectra are reproduced with a power-law model.
\par
\begin{longrotatetable}
\begin{deluxetable*}{lccccccccc}
\tablenum{2}
\tablecaption{Results of spectral fitting for the I\,Zw\,18 ULX\label{tab:bestfitparam}}
\tablewidth{0pt}
\tabletypesize{\footnotesize}
\tablehead{
\colhead{Model} & \colhead{$N_{\rm H}$} & \colhead{${\Gamma}$} & \colhead{$kT_{\rm in}$} & \colhead{$p$} &\colhead{$Norm._{\rm disk}$} &\colhead{$Norm._{\rm Comp}$} & \colhead{$f_{\rm 0.3-10\,keV}$} & \colhead{$L_{\rm 0.3-10\,keV}$} & \colhead{$\chi^{2}/d.o.f.$}\\
\colhead{} & \colhead{[${\rm 10^{21}\,cm^{-2}}$]} & \colhead{} & \colhead{[${\rm keV}$]} & \colhead{} & \colhead{} & \colhead{[${\rm keV^{-1} cm^{-2} s^{-1}}$]} & \colhead{$[{10^{-13}\,\rm erg\,s^{-1}\,cm^{-2}}]$} & \colhead{$[{10^{39}\,\rm erg\,s^{-1}}]$} & \colhead{}
}
\decimalcolnumbers
\startdata
\multicolumn{10}{c}{\it Chandra {\rm (2000-02-08)}}\\
\hline
{\tt diskbb} & $<0.88$ & $-$ & $0.91^{+0.15}_{-0.14}$ & $-$ & $4.6^{+3.5}_{-1.8}\times10^{-3}$ & $-$ & $0.62$ & $2.5$ & $26.5/21$ \\
{\tt powerlaw} & $<7.0$ & $1.9^{+0.2}_{-0.2}$ & $-$ & $-$ & $-$ & $1.5^{+0.2}_{-0.2}\times10^{-5}$ & $0.94$ & $3.7$ & $9.40/21$ \\
\hline
\multicolumn{10}{c}{\it XMM\,{\rm1} {\rm (2002-04-10)}}\\
\hline
{\tt diskbb} & $2.7^{+1.0}_{-0.8}$ & $-$ & $1.1^{+0.0}_{-0.1}$ & $-$ & $1.0^{+0.2}_{-0.2}\times10^{-2}$ & $-$ & $2.8$ & $11$ & $195.7/193$ \\
{\tt powerlaw} & $13^{+3}_{-1}$ & $2.0^{+0.1}_{-0.1}$ & $-$ & $-$ & $-$ & $7.1^{+0.5}_{-0.4}\times10^{-5}$ & $4.0$ & $16$ & $249.6/193$ \\
{\tt diskbb+powerlaw} & $5.4^{+2.9}_{-2.4}$ & $1.7^{+0.7}_{-1.0}$ & $1.0^{+0.2}_{-0.1}$ & $-$ & $9.4^{+7.8}_{-4.4}\times10^{-3}$  & $1.8^{+1.2}_{-1.5}\times10^{-5}$  & $3.1$ & $12$ &  180.5/191\\
{\tt diskpbb} & $6.0^{+2.1}_{-1.8}$ & $-$ & $1.3^{+0.2}_{-0.1}$ & $0.62^{+0.05}_{-0.04}$ & $2.4^{+2.4}_{-1.2}\times10^{-3}$  & $-$ & $3.1$ & 12 &  182.6/192\\
\hline
\multicolumn{10}{c}{\it XMM\,{\rm2}  {\rm (2002-04-16)}}\\
\hline
{\tt diskbb} & $<4.0$ & $-$ & $1.2^{+0.1}_{-0.1}$ & $-$ & $7.3^{+2.9}_{-2.1}\times10^{-3}$ & $-$ & $2.6$ & $11$ & $58.3/66$ \\
{\tt powerlaw} & $15^{+5}_{-4}$ & $1.9^{+0.2}_{-0.1}$ & $-$ & $-$ & $-$ & $6.7^{+0.9}_{-0.7}\times10^{-5}$ & $4.0$ & $16$ & $50.7/66$ \\
{\tt diskbb+powerlaw} & $8.2^{+9.1}_{-6.9}$ & $<3.5$ & $0.83^{+1.10}_{-0.28}$ & $-$ & $1.0^{+3.8}_{-1.0}\times10^{-2}$ & $3.4^{+3.1}_{-3.4}\times10^{-5}$ & $3.4$ & $13$ & $47.3/64$ \\
{\tt diskpbb} & $10^{+7}_{-5}$ & $-$ & $1.9^{+5.1}_{-0.5}$ & $0.56^{+0.07}_{-0.05}$ & $<2.0\times10^{-3}$ & $-$ & $<3.4$ & $<13$ & $48.0/65$ \\
\hline
\multicolumn{10}{c}{\it Suzaku\,{\rm1} {\rm (2014-05-15)}}\\
\hline
{\tt diskbb} & $<47$ & $-$ & $0.87^{+0.22}_{-0.19}$ & $-$ & $1.9^{+3.7}_{-1.2}\times10^{-2}$ & $-$ & $2.2$ & $8.6$ & $12.8/15$ \\
{\tt powerlaw} & $61^{+47}_{-31}$ & $2.7^{+0.5}_{-0.5}$ & $-$ & $-$ & $-$ & $1.1^{+0.8}_{-0.4}\times10^{-4}$ & $5.3$ & $21$ & $13.0/15$ \\
{\tt diskpbb} & $<73$ & $-$ & $1.1^{+1.0}_{-0.4}$ & (not constrained) & $2.4^{+52.5}_{-2.3}\times10^{-3}$ & $-$ & $3.1$ & $12$ & $12.5/14$ \\
\hline
\multicolumn{10}{c}{\it Suzaku\,{\rm2} {\rm (2014-10-04)}}\\
\hline
{\tt diskbb} & $<8.4$ & $-$ & $0.70^{+0.08}_{-0.10}$ & $-$ & $2.2^{+2.1}_{-0.8}\times10^{-2}$ & $-$ & $1.1$ & $4.3$ & $77.7/72$ \\
{\tt powerlaw} & $30^{+16}_{-13}$ & $2.9^{+0.3}_{-0.3}$ & $-$ & $-$ & $-$ & $5.7^{+1.7}_{-1.3}\times10^{-5}$ & $2.9$ & $11$ & $77.0/72$ \\
{\tt diskbb+powerlaw} & $<14$ & $1.7$ (fixed) & $0.63^{+0.13}_{-0.16}$ & $-$ & $3.0^{+6.7}_{-1.5}\times10^{-2}$ & $<9.4\times10^{-6}$ & $1.4$ & $5.4$ & $76.6/71$ \\
{\tt diskpbb} & $<23$ & $-$ & $0.96^{+0.28}_{-0.27}$ & $<0.73$ & $1.9^{+19.7}_{-1.3}\times10^{-3}$ & $-$ & $1.6$ & $6.5$ & $74.9/71$ \\
\enddata
\tablecomments{Best-fit parameters for each model with errors in the 90\,\% confidence level in 5 epochs. \\
(1) Model name in XSPEC.
(2) Specific absorption column density in the ULX. 
(3) Photon index of the Comptonization component. 
(4) Inner-disk temperature of the disk component. 
(5) Index of the disk radial dependence of  temperature, $T(r)\propto r^{-p}$. 
(6) Normalization  for the disk component, as $r^{2}_{\rm in}\,{\rm cos}\,i/{D}_{10}^2$.
(7) Normalization  for the Comptonization component, as photons ${\rm keV^{-1} cm^{-2} s^{-1}}$ at 1\,keV.
(8) Model flux in the energy range  ${\rm 0.3-10\,keV}$.
(9) Luminosity calculated multiplication of the ${\rm 0.3-10\,keV}$ model flux and  $4\pi\times(\mathrm{distance})^{2}$.
(10) Chi-square value of  model fitting and the degree of freedom.
}
\end{deluxetable*}
\end{longrotatetable}
We apply a few models to model fitting of the I\,Zw\,18 ULX spectra. First,  we apply the simple MCD model \citep[{\tt TBabs*TBvarabs*diskbb} in XSPEC;][]{1984PASJ...36..741M,1986ApJ...308..635M}. The model is composed of two parameters: an inner-disk temperature $kT_{\rm in}$ [keV] and a normalization,  $r^{2}_{\rm in}\,{\rm cos}\,i/{D}_{10}^2$, where $r_{\rm in}$ [km] is the apparent inner-disk radius, $i$ is the inclination angle, and $D_{10}$ [10 kpc] is the distance to the object. 
We find that the MCD model is rejected for none of the spectra at a significance level of 1\,\%.
The inner-disk temperature shows time variation;
the highest is $1.2^{+0.1}_{-0.1}\,{\rm keV}$ in epoch X2 and the lowest,  $0.70^{+0.08}_{-0.10}\,{\rm keV}$ in epoch S2.
The determined unabsorbed luminosity for any of the spectra exceeds the ULX  threshold of $10^{39}\,{\rm erg\,s^{-1}}$ in the energy range of $0.3-10\,{\rm keV}$. 
The unabsorbed luminosity increased by a factor of about four,  from $2.5\times10^{39}\,{\rm erg\,s^{-1}}$ in 2000 ({\it Chandra} observation) to $11\times10^{39}\,{\rm erg\,s^{-1}}$  in 2002 ({\it XMM-Newton} observations), and thus we confirm the previously-reported results \citep{2013ApJ...770...20K}. 
The observations in epochs S1 and S2 in 2014 show luminosities of $8.6\times10^{39}\,{\rm erg\,s^{-1}}$ and $4.3\times10^{39}\,{\rm erg\,s^{-1}}$, respectively, both of which are lower than those in the {\it XMM-Newton} observations in 2002.
\par
\citet{2013ApJ...770...20K} argued that the increase in the ULX luminosity between 2000 and 2002  was due to a hard-to-soft state transition of the ULX source.
Indeed,  fitting the data with a simple power-law model ({\tt TBabs*TBvarabs*powerlaw} in XSPEC) shows that the model is rejected for only the data in epoch {\it XMM}\,1, which has the highest data quality among our data sets,  at a significance level of 1\,\%. 
By contrast, the chi-square value of the ${\it Chandra}$ data is improved with the power-law model (${\chi^{2}/d.o.f}=9.40/21$)  from the MCD model (${\chi^{2}/d.o.f}=26.5/21$).
Although the MCD model is not rejected at a significance level of 1\,\% with the ${\it Chandra}$ data, the data show distinct systematic residuals (Figure \ref{fig:spectra_prev}).
Thus, our results support the hypothesis of the ULX state transition between 2000 and 2002. 
\par
The spectra of most ULXs are known to be reproducible with dual-continuum models that combine two of the following models: a blackbody, a MCD, a MCD with power-law dependence for the temperature \citep[{\tt diskpbb} in XSPEC; e.g.,][]{1994ApJ...426..308M}, a (cutoff) power-law, and a thermal Comptonization component \citep[e.g.,][]{2015MNRAS.454.3134M, 2017ApJ...836..113P, 2017A&A...608A..47K, 2018ApJ...856..128W}. 
We then test the dual-continuum model with the I\,Zw\,18 ULX data. A simple MCD + power-law model ({\tt TBabs*TBvarabs*(diskbb+powerlaw)} in XSPEC) is employed here because of the insufficient photon statistics, especially in the hard band ($> 2\,{\rm keV}$).
The parameters are found to be meaningfully constrained only with the good-statistics data of epochs X1, X2, and S2.
We note that in the fitting for epoch S2 data, the photon index of the power-law model is fixed at the best-fit value of the fitting of the X1 data.
The inner-disk temperatures are obtained to be $1.0^{+0.2}_{-0.1}$, $0.83^{+1.10}_{-0.28}$, and $0.63^{+0.13}_{-0.16}\,{\rm keV}$ for epoch X1, X2, and S2 data, respectively.
These values are almost the same as those determined with the single MCD model fitting for each observation.
This result suggests that the spectra are not dominated by the power-law component.
\par
We next explore a disk radial dependence of temperature $T(r)\propto r^{-p}$, where $p$ is a free parameter in the {\tt diskpbb} model ({\tt TBabs*TBvarabs*diskpbb} in XSPEC). 
We obtain  $kT_{\rm in}=1.3^{+0.2}_{-0.1}, 1.9^{+5.1}_{-0.5}$, and $0.96^{+0.28}_{-0.27}\,{\rm keV}$ and $p=0.62^{+0.05}_{-0.04}$, and $0.56^{+0.07}_{-0.05}, <0.73$ for epoch X1, X2, and S2 data, respectively. 
The obtained temperature $kT_{\rm in}$ is somewhat higher than those based on the other models.
For the {\it Suzaku}\,1 data, $kT_{\rm in}$ is poorly constrained, $1.1^{+1.0}_{-0.4}\,{\rm keV}$,  and $p$ is not constrained.
The {\it Chandra} data have inadequate statistics to yield significant results.
%
\section{Discussion} \label{sec:discussion}
\begin{figure}[th]
\epsscale{1.15} 
\plotone{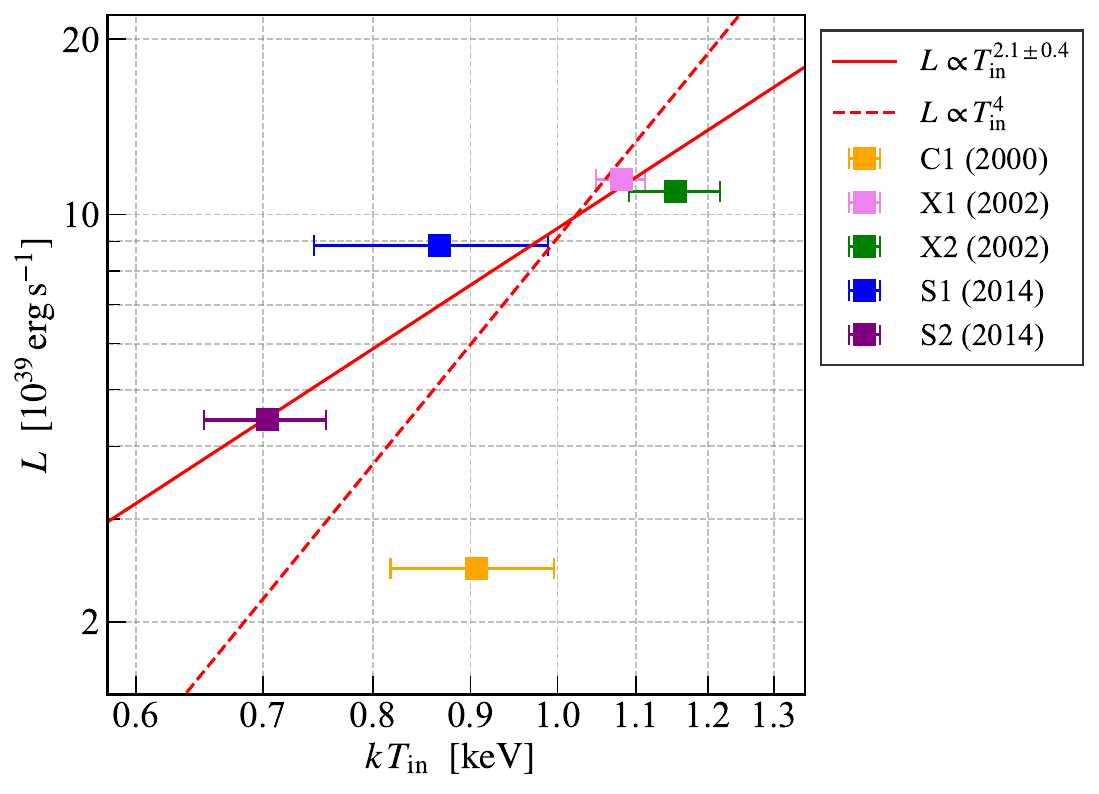}
\caption{ 
 Relation between the bolometric luminosity $L$ and inner-disk\ temperature $kT_{\rm in}$.  Error bars show the $1\,\sigma$ confidence intervals derived with the MCD model. 
The solid red line shows the best-fit relation of $L\propto T_{\rm in}^{2.1\pm0.4}$, where the \textit{Chandra} data are excluded from the fitting (see text for detail).  The dashed red line shows the nominal relation of $L\propto T_{\rm in}^{4}$  for the standard disk \citep{1973A&A....24..337S}.
\label{fig:L-T_relation}
}
\end{figure}

We discuss the time variation of the inner-disk temperature $kT_{\rm in}$  with the MCD model. 
The best-fit value of $kT_{\rm in}$ is in  a range of $0.7-1.2\,{\rm keV}$.
We here investigate a temperature dependence on the bolometric luminosity through a long-term spectral evolution.
For the standard disk---optically thick and geometrically thin---where a half of the gravitational potential energy is converted into blackbody radiation, the luminosity follows $L_{\rm bol}\propto T_{\rm in}^{4}$ \citep{1973A&A....24..337S}.
As the accretion rate increases, the slim disk---optically and geometrically thick---appears, where advection decreases the radiation efficiency, and the luminosity relation approaches $L_{\rm bol}\propto T_{\rm in}^{2}$ \citep{1988ApJ...332..646A, 2000PASJ...52..133W,2001PASJ...53..915W}.
Figure~\ref{fig:L-T_relation} plots the obtained data points on the $L-kT_{\rm in}$ space,  where the luminosity in an energy range of $0.1-50\,{\rm keV}$ is regarded as the bolometric luminosity $L$. 
We discuss the relation with $\chi^2$ test.
Using all the data, the luminosity relation model for the standard disk, $L\propto T_{\rm in}^{4}$,  is rejected at a significance level of 1\,\%. 
Also, that for the slim disk of $L\propto T_{\rm in}^{2}$ is rejected at a significance level of 1\,\%. 
The slim disk is expected to appear at a high Eddington ratio according to the standard model.
For this reason, we then exclude the {\it Chandra} data, which has the lowest luminosity in our data sets and shows apparent systematic residuals in our fitting result with the  \texttt{diskbb} model, and re-evaluate the above-mentioned relations.
 The evaluation shows that at a significance level of 1.5\,\%, $L\propto T_{\rm in}^{4}$ is rejected, whereas $L\propto T_{\rm in}^{2}$ is not.
When the power-law index in the relation is allowed to vary, the relation is derived to be $L\propto T_{\rm in}^{2.1\pm0.4}$, which is consistent with the expected property of a slim disk with supercritical accretion.
If the {\it Chandra} data is included in the fitting, the best-fit relation is $L\propto T_{\rm in}^{3.2\pm0.6}$, but the fitting is rejected at a significance level of 1\,\%.
Here, these errors are evaluated with $1\,\sigma$ confidence intervals.
These results suggest the hypothesis that most of the observations the data of which we have used in this work were performed when the ULX was in the slim disk state, and the {\it Chandra} observation was made when the ULX was in a different state from the others.
The nature of the \IZw ULX may be a stellar-mass compact object.
\par
\begin{figure}[th]
\epsscale{1.15} 
\plotone{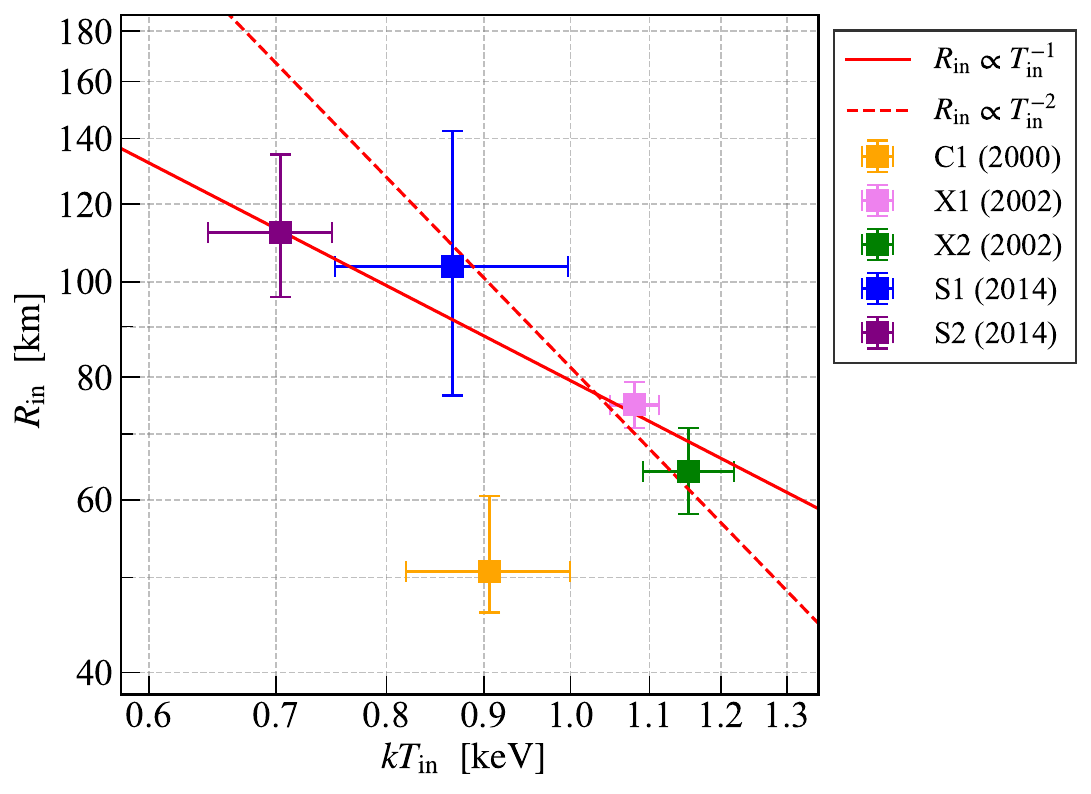}
\caption{ 
The apparent inner-disk radius $R_{\rm in}$ versus its temperature $kT_{\rm in}$ with $1\,\sigma$ confidence level.
The {\it Chandra} data point seems to be an outlier.
The radius $R_{\rm in}$ is not constant at a 7\,\% significance level,  where {\it Chandra} data point is excluded in the calculation.
It appears to be consistent with the relations $R_{\rm in}\propto T_{\rm in}^{-1}$ (solid red line) or $R_{\rm in}\propto T_{\rm in}^{-2}$ (dashed red line) for supercritical accretion.
\label{fig:R-T_relation}
}
\end{figure}

We also investigate the relation between the apparent inner-disk radius and the inner-disk temperature.
It should be noted that observed disk-blackbody spectra usually deviate from the MCD model due to inverse Compton scattering and a boundary condition at the innermost stable circular orbit (ISCO). 
Typically, the corrected apparent radius $R_{\rm in}$ for a standard disk is calculated  according to $\xi\kappa^2r_{\rm in} $, using a boundary-condition correction factor of $\xi=\sqrt{\frac{3}{7}}{(\frac{6}{7})^3}\sim0.41$ \citep{1998PASJ...50..667K} and a hardening factor for color-temperature correction of $\kappa\sim1.7$ \citep{1995ApJ...445..780S}. 
In Figure \ref{fig:R-T_relation}, we show the relation between the apparent radius $R_{\rm in}$ and the inner-disk temperature $kT_{\rm in}$ with our data, where we do not take into account the correction factor $\kappa$ in our calculation because in the low metallicity environment, the effect of electron scattering \citep[opacity and Comptonization;][]{2003ApJ...593...69K} deviates from the typical value.  
The apparent radius appears to decrease as the temperature increases with our data.
The {\it Chandra} data point appears to show a different trend from the other data.
For the standard disk, an apparent radius is expected to be constant in a thermal state.
When we fit the apparent radius with a constant model excluding the {\it Chandra} data, the model is rejected at a significance level of 7\,\%, reinforcing the negative relation.
The negative relation between $R_{\rm in}$ and $T_{\rm in}$ for ULXs were reported in several studies, $R_{\rm in}\propto T_{\rm in}^{-1}$ \citep{2000PASJ...52..133W,2001ApJ...554.1282M} or $R_{\rm in}\propto T_{\rm in}^{-2}$ \citep{2016MNRAS.456.1859U,2016MNRAS.456.1837S}, for supercritical accretion.
We overlay the two relations in Figure \ref{fig:R-T_relation}. 
The set of data of the \IZw ULX appear to match well the negative $R_{\rm in}-T_{\rm in}$ relation.
This result is consistent with the ULX being driven by supercritical accretion with a stellar-mass compact object.
\par
The strongest observational evidence for supercritical accretion is detection of emission and absorption lines originating in a fast outflow. They are highly redshifted or blueshifted in several ULXs \citep{2016Natur.533...64P,2017MNRAS.468.2865P}.
Because the photon statistics are not enough, we see no clear emission or absorption lines in our spectra.
\citet{2009MNRAS.398.1450K} classified 11 ULXs into three spectral types;  the power-law dominant type,  the MCD dominant type, which shows a positive $L-T_{\rm in}$ correlation, and the cool MCD ($0.1-0.3\,{\rm keV}$) + power-law type, which shows a negative correlation of $L\propto T_{\rm in}^{-3.5}$. 
In the negatively correlated case ``outflow-dominated ULXs'', the X-ray photons are  believed to originate from the photosphere of the optically thick outflow (funnel), whose temperature  decreases  as the mass accretion rate increases 
\citep{2007MNRAS.377.1187P}.
Indeed, a negative trend was found in NGC 1313 X-2, which is a pulsating neutron star in supercritical accretion \citep{2007ApJ...660L.113F, 2009MNRAS.398.1450K, 2021A&A...652A.118R}.
In our study, the relation between the luminosity and temperature is derived to be $L\propto T^{1.7\pm0.7}_{\rm in}$ $(1\,\sigma$ confidence interval) from MCD + power-law model fitting. 
The inner-disk temperature of the \IZw ULX is higher ($>0.5\,{\rm keV}$) than those of outflow-dominated ULXs. Therefore, we consider that the accretion disk is more predominant than the outflow in the \IZw ULX. 
\par
The spectral model fitting with the {\tt diskpbb} model is an effective way to distinguish the disk states. The standard disk is reproduced with $p=0.75$ \citep{1973A&A....24..337S}, whereas the $p$-value for the slim disk appears to approach $p=0.50$ \citep{1999PASJ...51..725W,2000PASJ...52..133W}.
For the \IZw ULX, the disk radial dependence of temperature is $p<0.75$ for $T(r)\propto r^{-p}$.
The value is significantly different from that for the standard disk. 
This result also reinforces the hypothesis that \IZw ULX is in supercritical accretion onto a stellar-mass compact object.  
\par
\IZw is thought to have begun star formation more than 1\,Gyr ago \citep{2007ApJ...667L.151A}.
According to \citet{2017ApJ...846...17W}, if a SFR has been constant since the beginning of star formation in a low-metallicity environment, ULXs are likely to be BHs.
On the contrary, if a burst star-forming occurred in the past, ULXs could be either BHs or NSs.
With the limited photon statistics of the currently available X-ray observation data for the \IZw ULX, we cannot determine whether it is a BH or NS. 
In addition to spectral analysis,  search for pulsation, quasi-periodic oscillation, or flat-topped noise \citep[e.g.,][]{2019MNRAS.486.2766A} with longer observations in the future will be useful to determine what the \IZw ULX actually is.
%
\section{Conclusions} \label{sec:conclusions}
In this study, we investigate the spectral evolution of the ULX in metal-poor galaxy \IZw for fourteen years, using five observations from {\it Chandra}, {\it XMM$-$Newton}, and {\it Suzaku}.
The inclusion of the data from two previously-unpublished {\it Suzaku}  observations allows us to examine spectral parameter correlations. It also helps us to confirm that the {\it Chandra}  observation was made when the source was in a different state from other epochs.
We suggest that the \IZw ULX is driven with a supercritical accreting flow with a stellar-mass compact object based on the following findings:
\begin{enumerate}
\item  The \IZw ULX shows a positive correlation between the bolometric luminosity $L$ and the inner-disk temperature $T_{\rm in}$, which we derive to be $L\propto T_{\rm in}^{2.1\pm0.4}$  with the MCD model, where we exclude the \textit{Chandra} data because the source state differed at the time of the \textit{Chandra} observation.
The relation $L\propto T_{\rm in}^{4}$ for the standard disk is rejected at a significance level of 1.5\,\%.
\item The apparent inner-disk radius $R_{\rm in}$ is not constant  at a significance level of 7\,\%, where the \textit{Chandra} data are excluded in the estimate.
The radius appears negatively correlated with the disk temperature in a similar way as previously suggested for some other ULXs with $R_{\rm in}\propto T_{\rm in}^{-1}$ or $R_{\rm in}\propto T_{\rm in}^{-2}$.
\item The radial dependence of the disk temperature is obtained to be $T (r)\propto r^{-p}$ with $p<0.75$.  This is consistent with the slim disk ($p\sim0.5$) and differs from the standard disk ($p=0.75$).
\end{enumerate}
\par
In general, low metallicity environments are considered to facilitate the formation of heavy BHs. 
It has been believed that IMBH ULXs are abundant in such environments.
However, the ULX in metal-poor galaxy \IZw is a possible stellar-mass compact object.
To determine the general nature of the ULX  is essential for delving into the evolution of binary systems.
Longer observations and systematic studies are needed to reveal what factors determine their nature.
\newpage
\begin{acknowledgments}
\par
We would like to thank the anonymous referee for his/her valuable comments and suggestions.
This work was conducted with the late Dr. K. Hayashida. 
This research has made use of data obtained from the High Energy Astrophysics Science Archive Research Center (HEASARC), which is a service of the Astrophysics Science Division at NASA/GSFC, and the XMM-Newton Science Archive (XSA). 
This research also employs a Chandra dataset, obtained by the Chandra X-ray Observatory, contained in the Chandra Data Collection (CDC) 239~\dataset[doi:10.25574/cdc.239]{https://doi.org/10.25574/cdc.239}.
The data of {\it Suzaku} satellite, which is a collaborative mission between the space agencies of Japan (JAXA) and the NASA, the USA played an important role in this research. 
We use X-ray data analysis software of the application packages HEASoft, CIAO, and SAS provided by the HEASARC, the Chandra X-ray Center (CXC), and the ESA XMM Science Operations Centre (SOC), respectively.
This work is supported by the Japan Science and Technology Agency (JST) Support for Pioneering Research Initiated by the Next Generation (SPRING) with Grant Number JPMJSP2138, the Japan Society for the Promotion of Science (JSPS) Grants-in-Aid for Scientific Research (KAKENHI) with Grant Numbers 19K21884, 20H00175, 20H00178, 20H01941, 20H01947, 20KK0071, 21K20372, 22H00128, 22K18277, and 23H00128, and the Toray Science and Technology Grant Number 20-6104 (Toray Science Foundation).
\end{acknowledgments}
\par
\newpage
\vspace{5mm}
\facilities{CXO (ACIS), XMM (EPIC), Suzaku (XIS)}
\software{HEAsoft\,\citep{2014ascl.soft08004N}, CIAO\,\citep{2006SPIE.6270E..1VF}, SAS, XSPEC\,\citep{1996ASPC..101...17A}, SAOImageDS9\,\citep{2003ASPC..295..489J}, numpy\,\citep{harris2020array}, scipy\,\citep{2020SciPy-NMeth}, matplotlib\,\citep{Hunter:2007}, astropy\,\citep{2013A&A...558A..33A,2018AJ....156..123A,2022ApJ...935..167A}
}

\bibliography{20240520_IZw18}{}

\begin{thebibliography}{}
\expandafter\ifx\csname natexlab\endcsname\relax\def\natexlab#1{#1}\fi
\providecommand{\url}[1]{\href{#1}{#1}}
\providecommand{\dodoi}[1]{doi:~\href{http://doi.org/#1}{\nolinkurl{#1}}}
\providecommand{\doeprint}[1]{\href{http://ascl.net/#1}{\nolinkurl{http://ascl.net/#1}}}
\providecommand{\doarXiv}[1]{\href{https://arxiv.org/abs/#1}{\nolinkurl{https://arxiv.org/abs/#1}}}

\bibitem[{{Abramowicz} {et~al.}(1988){Abramowicz}, {Czerny}, {Lasota}, \&
  {Szuszkiewicz}}]{1988ApJ...332..646A}
{Abramowicz}, M.~A., {Czerny}, B., {Lasota}, J.~P., \& {Szuszkiewicz}, E. 1988,
  \apj, 332, 646, \dodoi{10.1086/166683}

\bibitem[{{Akylas} {et~al.}(2006){Akylas}, {Georgantopoulos}, {Georgakakis},
  {Kitsionas}, \& {Hatziminaoglou}}]{2006A&A...459..693A}
{Akylas}, A., {Georgantopoulos}, I., {Georgakakis}, A., {Kitsionas}, S., \&
  {Hatziminaoglou}, E. 2006, \aap, 459, 693, \dodoi{10.1051/0004-6361:20054632}

\bibitem[{{Aloisi} {et~al.}(2007){Aloisi}, {Clementini}, {Tosi}, {Annibali},
  {Contreras}, {Fiorentino}, {Mack}, {Marconi}, {Musella}, {Saha}, {Sirianni},
  \& {van der Marel}}]{2007ApJ...667L.151A}
{Aloisi}, A., {Clementini}, G., {Tosi}, M., {et~al.} 2007, \apjl, 667, L151,
  \dodoi{10.1086/522368}

\bibitem[{{Arnaud}(1996)}]{1996ASPC..101...17A}
{Arnaud}, K.~A. 1996, in Astronomical Society of the Pacific Conference Series,
  Vol. 101, Astronomical Data Analysis Software and Systems V, ed. G.~H.
  {Jacoby} \& J.~{Barnes}, 17

\bibitem[{{Asplund} {et~al.}(2009){Asplund}, {Grevesse}, {Sauval}, \&
  {Scott}}]{2009ARA&A..47..481A}
{Asplund}, M., {Grevesse}, N., {Sauval}, A.~J., \& {Scott}, P. 2009, \araa, 47,
  481, \dodoi{10.1146/annurev.astro.46.060407.145222}

\bibitem[{{Astropy Collaboration} {et~al.}(2013){Astropy Collaboration},
  {Robitaille}, {Tollerud}, {Greenfield}, {Droettboom}, {Bray}, {Aldcroft},
  {Davis}, {Ginsburg}, {Price-Whelan}, {Kerzendorf}, {Conley}, {Crighton},
  {Barbary}, {Muna}, {Ferguson}, {Grollier}, {Parikh}, {Nair}, {Unther},
  {Deil}, {Woillez}, {Conseil}, {Kramer}, {Turner}, {Singer}, {Fox}, {Weaver},
  {Zabalza}, {Edwards}, {Azalee Bostroem}, {Burke}, {Casey}, {Crawford},
  {Dencheva}, {Ely}, {Jenness}, {Labrie}, {Lim}, {Pierfederici}, {Pontzen},
  {Ptak}, {Refsdal}, {Servillat}, \& {Streicher}}]{2013A&A...558A..33A}
{Astropy Collaboration}, {Robitaille}, T.~P., {Tollerud}, E.~J., {et~al.} 2013,
  \aap, 558, A33, \dodoi{10.1051/0004-6361/201322068}

\bibitem[{{Astropy Collaboration} {et~al.}(2018){Astropy Collaboration},
  {Price-Whelan}, {Sip{\H{o}}cz}, {G{\"u}nther}, {Lim}, {Crawford}, {Conseil},
  {Shupe}, {Craig}, {Dencheva}, {Ginsburg}, {VanderPlas}, {Bradley},
  {P{\'e}rez-Su{\'a}rez}, {de Val-Borro}, {Aldcroft}, {Cruz}, {Robitaille},
  {Tollerud}, {Ardelean}, {Babej}, {Bach}, {Bachetti}, {Bakanov}, {Bamford},
  {Barentsen}, {Barmby}, {Baumbach}, {Berry}, {Biscani}, {Boquien}, {Bostroem},
  {Bouma}, {Brammer}, {Bray}, {Breytenbach}, {Buddelmeijer}, {Burke},
  {Calderone}, {Cano Rodr{\'\i}guez}, {Cara}, {Cardoso}, {Cheedella}, {Copin},
  {Corrales}, {Crichton}, {D'Avella}, {Deil}, {Depagne}, {Dietrich}, {Donath},
  {Droettboom}, {Earl}, {Erben}, {Fabbro}, {Ferreira}, {Finethy}, {Fox},
  {Garrison}, {Gibbons}, {Goldstein}, {Gommers}, {Greco}, {Greenfield},
  {Groener}, {Grollier}, {Hagen}, {Hirst}, {Homeier}, {Horton}, {Hosseinzadeh},
  {Hu}, {Hunkeler}, {Ivezi{\'c}}, {Jain}, {Jenness}, {Kanarek}, {Kendrew},
  {Kern}, {Kerzendorf}, {Khvalko}, {King}, {Kirkby}, {Kulkarni}, {Kumar},
  {Lee}, {Lenz}, {Littlefair}, {Ma}, {Macleod}, {Mastropietro}, {McCully},
  {Montagnac}, {Morris}, {Mueller}, {Mumford}, {Muna}, {Murphy}, {Nelson},
  {Nguyen}, {Ninan}, {N{\"o}the}, {Ogaz}, {Oh}, {Parejko}, {Parley}, {Pascual},
  {Patil}, {Patil}, {Plunkett}, {Prochaska}, {Rastogi}, {Reddy Janga},
  {Sabater}, {Sakurikar}, {Seifert}, {Sherbert}, {Sherwood-Taylor}, {Shih},
  {Sick}, {Silbiger}, {Singanamalla}, {Singer}, {Sladen}, {Sooley},
  {Sornarajah}, {Streicher}, {Teuben}, {Thomas}, {Tremblay}, {Turner},
  {Terr{\'o}n}, {van Kerkwijk}, {de la Vega}, {Watkins}, {Weaver}, {Whitmore},
  {Woillez}, {Zabalza}, \& {Astropy Contributors}}]{2018AJ....156..123A}
{Astropy Collaboration}, {Price-Whelan}, A.~M., {Sip{\H{o}}cz}, B.~M., {et~al.}
  2018, \aj, 156, 123, \dodoi{10.3847/1538-3881/aabc4f}

\bibitem[{{Astropy Collaboration} {et~al.}(2022){Astropy Collaboration},
  {Price-Whelan}, {Lim}, {Earl}, {Starkman}, {Bradley}, {Shupe}, {Patil},
  {Corrales}, {Brasseur}, {N{\"o}the}, {Donath}, {Tollerud}, {Morris},
  {Ginsburg}, {Vaher}, {Weaver}, {Tocknell}, {Jamieson}, {van Kerkwijk},
  {Robitaille}, {Merry}, {Bachetti}, {G{\"u}nther}, {Aldcroft},
  {Alvarado-Montes}, {Archibald}, {B{\'o}di}, {Bapat}, {Barentsen},
  {Baz{\'a}n}, {Biswas}, {Boquien}, {Burke}, {Cara}, {Cara}, {Conroy},
  {Conseil}, {Craig}, {Cross}, {Cruz}, {D'Eugenio}, {Dencheva}, {Devillepoix},
  {Dietrich}, {Eigenbrot}, {Erben}, {Ferreira}, {Foreman-Mackey}, {Fox},
  {Freij}, {Garg}, {Geda}, {Glattly}, {Gondhalekar}, {Gordon}, {Grant},
  {Greenfield}, {Groener}, {Guest}, {Gurovich}, {Handberg}, {Hart},
  {Hatfield-Dodds}, {Homeier}, {Hosseinzadeh}, {Jenness}, {Jones}, {Joseph},
  {Kalmbach}, {Karamehmetoglu}, {Ka{\l}uszy{\'n}ski}, {Kelley}, {Kern},
  {Kerzendorf}, {Koch}, {Kulumani}, {Lee}, {Ly}, {Ma}, {MacBride}, {Maljaars},
  {Muna}, {Murphy}, {Norman}, {O'Steen}, {Oman}, {Pacifici}, {Pascual},
  {Pascual-Granado}, {Patil}, {Perren}, {Pickering}, {Rastogi}, {Roulston},
  {Ryan}, {Rykoff}, {Sabater}, {Sakurikar}, {Salgado}, {Sanghi}, {Saunders},
  {Savchenko}, {Schwardt}, {Seifert-Eckert}, {Shih}, {Jain}, {Shukla}, {Sick},
  {Simpson}, {Singanamalla}, {Singer}, {Singhal}, {Sinha}, {Sip{\H{o}}cz},
  {Spitler}, {Stansby}, {Streicher}, {{\v{S}}umak}, {Swinbank}, {Taranu},
  {Tewary}, {Tremblay}, {de Val-Borro}, {Van Kooten}, {Vasovi{\'c}}, {Verma},
  {de Miranda Cardoso}, {Williams}, {Wilson}, {Winkel}, {Wood-Vasey}, {Xue},
  {Yoachim}, {Zhang}, {Zonca}, \& {Astropy Project
  Contributors}}]{2022ApJ...935..167A}
{Astropy Collaboration}, {Price-Whelan}, A.~M., {Lim}, P.~L., {et~al.} 2022,
  \apj, 935, 167, \dodoi{10.3847/1538-4357/ac7c74}

\bibitem[{{Atapin} {et~al.}(2019){Atapin}, {Fabrika}, \&
  {Caballero-Garc{\'\i}a}}]{2019MNRAS.486.2766A}
{Atapin}, K., {Fabrika}, S., \& {Caballero-Garc{\'\i}a}, M.~D. 2019, \mnras,
  486, 2766, \dodoi{10.1093/mnras/stz1027}

\bibitem[{{Bachetti}(2016)}]{2016AN....337..349B}
{Bachetti}, M. 2016, Astronomische Nachrichten, 337, 349,
  \dodoi{10.1002/asna.201612312}

\bibitem[{{Bachetti} {et~al.}(2013){Bachetti}, {Rana}, {Walton}, {Barret},
  {Harrison}, {Boggs}, {Christensen}, {Craig}, {Fabian}, {F{\"u}rst},
  {Grefenstette}, {Hailey}, {Hornschemeier}, {Madsen}, {Miller}, {Ptak},
  {Stern}, {Webb}, \& {Zhang}}]{2013ApJ...778..163B}
{Bachetti}, M., {Rana}, V., {Walton}, D.~J., {et~al.} 2013, \apj, 778, 163,
  \dodoi{10.1088/0004-637X/778/2/163}

\bibitem[{{Bachetti} {et~al.}(2014){Bachetti}, {Harrison}, {Walton},
  {Grefenstette}, {Chakrabarty}, {F{\"u}rst}, {Barret}, {Beloborodov}, {Boggs},
  {Christensen}, {Craig}, {Fabian}, {Hailey}, {Hornschemeier}, {Kaspi},
  {Kulkarni}, {Maccarone}, {Miller}, {Rana}, {Stern}, {Tendulkar}, {Tomsick},
  {Webb}, \& {Zhang}}]{2014Natur.514..202B}
{Bachetti}, M., {Harrison}, F.~A., {Walton}, D.~J., {et~al.} 2014, \nat, 514,
  202, \dodoi{10.1038/nature13791}

\bibitem[{{Belczynski} {et~al.}(2010){Belczynski}, {Bulik}, {Fryer}, {Ruiter},
  {Valsecchi}, {Vink}, \& {Hurley}}]{2010ApJ...714.1217B}
{Belczynski}, K., {Bulik}, T., {Fryer}, C.~L., {et~al.} 2010, \apj, 714, 1217,
  \dodoi{10.1088/0004-637X/714/2/1217}

\bibitem[{{Bomans} \& {Weis}(2002)}]{2002ASPC..262..141B}
{Bomans}, D.~J., \& {Weis}, K. 2002, in Astronomical Society of the Pacific
  Conference Series, Vol. 262, The High Energy Universe at Sharp Focus: Chandra
  Science, ed. E.~M. {Schlegel} \& S.~D. {Vrtilek}, 141

\bibitem[{{Colbert} \& {Mushotzky}(1999)}]{1999ApJ...519...89C}
{Colbert}, E. J.~M., \& {Mushotzky}, R.~F. 1999, \apj, 519, 89,
  \dodoi{10.1086/307356}

\bibitem[{{Evans} {et~al.}(2020){Evans}, {Primini}, {Miller}, {Evans}, {Allen},
  {Anderson}, {Becker}, {Budynkiewicz}, {Burke}, {Chen}, {Civano}, {D'Abrusco},
  {Doe}, {Fabbiano}, {Martinez Galarza}, {Gibbs}, {Glotfelty}, {Graessle},
  {Grier}, {Hain}, {Hall}, {Harbo}, {Houck}, {Lauer}, {Laurino}, {Lee},
  {McCollough}, {McDowell}, {McLaughlin}, {Morgan}, {Mossman}, {Nguyen},
  {Nichols}, {Nowak}, {Paxson}, {Perdikeas}, {Plummer}, {Rots},
  {Siemiginowska}, {Sundheim}, {Thong}, {Tibbetts}, {Van Stone}, {Winkelman},
  \& {Zografou}}]{2020AAS...23515405E}
{Evans}, I.~N., {Primini}, F.~A., {Miller}, J.~B., {et~al.} 2020, in American
  Astronomical Society Meeting Abstracts, Vol. 235, American Astronomical
  Society Meeting Abstracts \#235, 154.05

\bibitem[{{Fabbiano}(1989)}]{1989ARA&A..27...87F}
{Fabbiano}, G. 1989, \araa, 27, 87, \dodoi{10.1146/annurev.aa.27.090189.000511}

\bibitem[{{Feng} \& {Kaaret}(2007)}]{2007ApJ...660L.113F}
{Feng}, H., \& {Kaaret}, P. 2007, \apjl, 660, L113, \dodoi{10.1086/518309}

\bibitem[{{Feng} \& {Soria}(2011)}]{2011NewAR..55..166F}
{Feng}, H., \& {Soria}, R. 2011, \nar, 55, 166,
  \dodoi{10.1016/j.newar.2011.08.002}

\bibitem[{{Fiorentino} {et~al.}(2010){Fiorentino}, {Contreras Ramos},
  {Clementini}, {Marconi}, {Musella}, {Aloisi}, {Annibali}, {Saha}, {Tosi}, \&
  {van der Marel}}]{2010ApJ...711..808F}
{Fiorentino}, G., {Contreras Ramos}, R., {Clementini}, G., {et~al.} 2010, \apj,
  711, 808, \dodoi{10.1088/0004-637X/711/2/808}

\bibitem[{{Fourniol} {et~al.}(1996){Fourniol}, {Pakull}, \&
  {Motch}}]{1996rftu.proc..375F}
{Fourniol}, N., {Pakull}, M.~W., \& {Motch}, C. 1996, in Roentgenstrahlung from
  the Universe, ed. H.~U. {Zimmermann}, J.~{Tr{\"u}mper}, \& H.~{Yorke},
  375--376

\bibitem[{{Fruscione} {et~al.}(2006){Fruscione}, {McDowell}, {Allen},
  {Brickhouse}, {Burke}, {Davis}, {Durham}, {Elvis}, {Galle}, {Harris},
  {Huenemoerder}, {Houck}, {Ishibashi}, {Karovska}, {Nicastro}, {Noble},
  {Nowak}, {Primini}, {Siemiginowska}, {Smith}, \&
  {Wise}}]{2006SPIE.6270E..1VF}
{Fruscione}, A., {McDowell}, J.~C., {Allen}, G.~E., {et~al.} 2006, in Society
  of Photo-Optical Instrumentation Engineers (SPIE) Conference Series, Vol.
  6270, Society of Photo-Optical Instrumentation Engineers (SPIE) Conference
  Series, ed. D.~R. {Silva} \& R.~E. {Doxsey}, 62701V,
  \dodoi{10.1117/12.671760}

\bibitem[{Harris {et~al.}(2020)Harris, Millman, van~der Walt, Gommers,
  Virtanen, Cournapeau, Wieser, Taylor, Berg, Smith, Kern, Picus, Hoyer, van
  Kerkwijk, Brett, Haldane, del R{\'{i}}o, Wiebe, Peterson,
  G{\'{e}}rard-Marchant, Sheppard, Reddy, Weckesser, Abbasi, Gohlke, \&
  Oliphant}]{harris2020array}
Harris, C.~R., Millman, K.~J., van~der Walt, S.~J., {et~al.} 2020, Nature, 585,
  357, \dodoi{10.1038/s41586-020-2649-2}

\bibitem[{{HI4PI Collaboration} {et~al.}(2016){HI4PI Collaboration}, {Ben
  Bekhti}, {Fl{\"o}er}, {Keller}, {Kerp}, {Lenz}, {Winkel}, {Bailin},
  {Calabretta}, {Dedes}, {Ford}, {Gibson}, {Haud}, {Janowiecki}, {Kalberla},
  {Lockman}, {McClure-Griffiths}, {Murphy}, {Nakanishi}, {Pisano}, \&
  {Staveley-Smith}}]{2016A&A...594A.116H}
{HI4PI Collaboration}, {Ben Bekhti}, N., {Fl{\"o}er}, L., {et~al.} 2016, \aap,
  594, A116, \dodoi{10.1051/0004-6361/201629178}

\bibitem[{{Hirano} {et~al.}(2014){Hirano}, {Hosokawa}, {Yoshida}, {Umeda},
  {Omukai}, {Chiaki}, \& {Yorke}}]{2014ApJ...781...60H}
{Hirano}, S., {Hosokawa}, T., {Yoshida}, N., {et~al.} 2014, \apj, 781, 60,
  \dodoi{10.1088/0004-637X/781/2/60}

\bibitem[{{Hunt} {et~al.}(2003){Hunt}, {Thuan}, \&
  {Izotov}}]{2003ApJ...588..281H}
{Hunt}, L.~K., {Thuan}, T.~X., \& {Izotov}, Y.~I. 2003, \apj, 588, 281,
  \dodoi{10.1086/368352}

\bibitem[{Hunter(2007)}]{Hunter:2007}
Hunter, J.~D. 2007, Computing in Science \& Engineering, 9, 90,
  \dodoi{10.1109/MCSE.2007.55}

\bibitem[{{Irwin} {et~al.}(2004){Irwin}, {Bregman}, \&
  {Athey}}]{2004ApJ...601L.143I}
{Irwin}, J.~A., {Bregman}, J.~N., \& {Athey}, A.~E. 2004, \apjl, 601, L143,
  \dodoi{10.1086/382026}

\bibitem[{{Ishisaki} {et~al.}(2007){Ishisaki}, {Maeda}, {Fujimoto}, {Ozaki},
  {Ebisawa}, {Takahashi}, {Ueda}, {Ogasaka}, {Ptak}, {Mukai}, {Hamaguchi},
  {Hirayama}, {Kotani}, {Kubo}, {Shibata}, {Ebara}, {Furuzawa}, {Iizuka},
  {Inoue}, {Mori}, {Okada}, {Yokoyama}, {Matsumoto}, {Nakajima}, {Yamaguchi},
  {Anabuki}, {Tawa}, {Nagai}, {Katsuda}, {Hayashida}, {Bamba}, {Miller},
  {Sato}, \& {Yamasaki}}]{2007PASJ...59S.113I}
{Ishisaki}, Y., {Maeda}, Y., {Fujimoto}, R., {et~al.} 2007, \pasj, 59, 113,
  \dodoi{10.1093/pasj/59.sp1.S113}

\bibitem[{{Izotov} \& {Thuan}(2004)}]{2004ApJ...616..768I}
{Izotov}, Y.~I., \& {Thuan}, T.~X. 2004, \apj, 616, 768, \dodoi{10.1086/424990}

\bibitem[{{Joye} \& {Mandel}(2003)}]{2003ASPC..295..489J}
{Joye}, W.~A., \& {Mandel}, E. 2003, in Astronomical Society of the Pacific
  Conference Series, Vol. 295, Astronomical Data Analysis Software and Systems
  XII, ed. H.~E. {Payne}, R.~I. {Jedrzejewski}, \& R.~N. {Hook}, 489

\bibitem[{{Kaaret} \& {Feng}(2013)}]{2013ApJ...770...20K}
{Kaaret}, P., \& {Feng}, H. 2013, \apj, 770, 20,
  \dodoi{10.1088/0004-637X/770/1/20}

\bibitem[{{Kaaret} {et~al.}(2017){Kaaret}, {Feng}, \&
  {Roberts}}]{2017ARA&A..55..303K}
{Kaaret}, P., {Feng}, H., \& {Roberts}, T.~P. 2017, \araa, 55, 303,
  \dodoi{10.1146/annurev-astro-091916-055259}

\bibitem[{{Kaaret} {et~al.}(2011){Kaaret}, {Schmitt}, \&
  {Gorski}}]{2011ApJ...741...10K}
{Kaaret}, P., {Schmitt}, J., \& {Gorski}, M. 2011, \apj, 741, 10,
  \dodoi{10.1088/0004-637X/741/1/10}

\bibitem[{{Kajava} \& {Poutanen}(2009)}]{2009MNRAS.398.1450K}
{Kajava}, J. J.~E., \& {Poutanen}, J. 2009, \mnras, 398, 1450,
  \dodoi{10.1111/j.1365-2966.2009.15215.x}

\bibitem[{{Kawaguchi}(2003)}]{2003ApJ...593...69K}
{Kawaguchi}, T. 2003, \apj, 593, 69, \dodoi{10.1086/376404}

\bibitem[{{Kehrig} {et~al.}(2021){Kehrig}, {Guerrero}, {V{\'\i}lchez}, \&
  {Ramos-Larios}}]{2021ApJ...908L..54K}
{Kehrig}, C., {Guerrero}, M.~A., {V{\'\i}lchez}, J.~M., \& {Ramos-Larios}, G.
  2021, \apjl, 908, L54, \dodoi{10.3847/2041-8213/abe41b}

\bibitem[{{Kehrig} {et~al.}(2016){Kehrig}, {V{\'\i}lchez}, {P{\'e}rez-Montero},
  {Iglesias-P{\'a}ramo}, {Hern{\'a}ndez-Fern{\'a}ndez}, {Duarte Puertas},
  {Brinchmann}, {Durret}, \& {Kunth}}]{2016MNRAS.459.2992K}
{Kehrig}, C., {V{\'\i}lchez}, J.~M., {P{\'e}rez-Montero}, E., {et~al.} 2016,
  \mnras, 459, 2992, \dodoi{10.1093/mnras/stw806}

\bibitem[{{King} {et~al.}(2023){King}, {Lasota}, \&
  {Middleton}}]{2023NewAR..9601672K}
{King}, A., {Lasota}, J.-P., \& {Middleton}, M. 2023, \nar, 96, 101672,
  \dodoi{10.1016/j.newar.2022.101672}

\bibitem[{{Koliopanos} {et~al.}(2017){Koliopanos}, {Vasilopoulos}, {Godet},
  {Bachetti}, {Webb}, \& {Barret}}]{2017A&A...608A..47K}
{Koliopanos}, F., {Vasilopoulos}, G., {Godet}, O., {et~al.} 2017, \aap, 608,
  A47, \dodoi{10.1051/0004-6361/201730922}

\bibitem[{{Kubota} {et~al.}(1998){Kubota}, {Tanaka}, {Makishima}, {Ueda},
  {Dotani}, {Inoue}, \& {Yamaoka}}]{1998PASJ...50..667K}
{Kubota}, A., {Tanaka}, Y., {Makishima}, K., {et~al.} 1998, \pasj, 50, 667,
  \dodoi{10.1093/pasj/50.6.667}

\bibitem[{{Kunth} \& {{\"O}stlin}(2000)}]{2000A&ARv..10....1K}
{Kunth}, D., \& {{\"O}stlin}, G. 2000, \aapr, 10, 1,
  \dodoi{10.1007/s001590000005}

\bibitem[{{Lebouteiller} {et~al.}(2017){Lebouteiller}, {P{\'e}quignot},
  {Cormier}, {Madden}, {Pakull}, {Kunth}, {Galliano}, {Chevance}, {Heap},
  {Lee}, \& {Polles}}]{2017A&A...602A..45L}
{Lebouteiller}, V., {P{\'e}quignot}, D., {Cormier}, D., {et~al.} 2017, \aap,
  602, A45, \dodoi{10.1051/0004-6361/201629675}

\bibitem[{{Linden} {et~al.}(2010){Linden}, {Kalogera}, {Sepinsky}, {Prestwich},
  {Zezas}, \& {Gallagher}}]{2010ApJ...725.1984L}
{Linden}, T., {Kalogera}, V., {Sepinsky}, J.~F., {et~al.} 2010, \apj, 725,
  1984, \dodoi{10.1088/0004-637X/725/2/1984}

\bibitem[{{Makishima} {et~al.}(1986){Makishima}, {Maejima}, {Mitsuda}, {Bradt},
  {Remillard}, {Tuohy}, {Hoshi}, \& {Nakagawa}}]{1986ApJ...308..635M}
{Makishima}, K., {Maejima}, Y., {Mitsuda}, K., {et~al.} 1986, \apj, 308, 635,
  \dodoi{10.1086/164534}

\bibitem[{{Makishima} {et~al.}(2000){Makishima}, {Kubota}, {Mizuno}, {Ohnishi},
  {Tashiro}, {Aruga}, {Asai}, {Dotani}, {Mitsuda}, {Ueda}, {Uno}, {Yamaoka},
  {Ebisawa}, {Kohmura}, \& {Okada}}]{2000ApJ...535..632M}
{Makishima}, K., {Kubota}, A., {Mizuno}, T., {et~al.} 2000, \apj, 535, 632,
  \dodoi{10.1086/308868}

\bibitem[{{Mapelli} {et~al.}(2011){Mapelli}, {Ripamonti}, {Zampieri}, \&
  {Colpi}}]{2011AN....332..414M}
{Mapelli}, M., {Ripamonti}, E., {Zampieri}, L., \& {Colpi}, M. 2011,
  Astronomische Nachrichten, 332, 414, \dodoi{10.1002/asna.201011511}

\bibitem[{{Mapelli} {et~al.}(2010){Mapelli}, {Ripamonti}, {Zampieri}, {Colpi},
  \& {Bressan}}]{2010MNRAS.408..234M}
{Mapelli}, M., {Ripamonti}, E., {Zampieri}, L., {Colpi}, M., \& {Bressan}, A.
  2010, \mnras, 408, 234, \dodoi{10.1111/j.1365-2966.2010.17048.x}

\bibitem[{{Marconi} {et~al.}(2010){Marconi}, {Musella}, {Fiorentino},
  {Clementini}, {Aloisi}, {Annibali}, {Contreras Ramos}, {Saha}, {Tosi}, \&
  {van der Marel}}]{2010ApJ...713..615M}
{Marconi}, M., {Musella}, I., {Fiorentino}, G., {et~al.} 2010, \apj, 713, 615,
  \dodoi{10.1088/0004-637X/713/1/615}

\bibitem[{{Martin}(1996)}]{1996ApJ...465..680M}
{Martin}, C.~L. 1996, \apj, 465, 680, \dodoi{10.1086/177453}

\bibitem[{{Matsumoto} {et~al.}(2003){Matsumoto}, {Tsuru}, {Watari},
  {Mineshige}, \& {Matsushita}}]{2003ASPC..289..291M}
{Matsumoto}, H., {Tsuru}, T.~G., {Watari}, K., {Mineshige}, S., \&
  {Matsushita}, S. 2003, in Astronomical Society of the Pacific Conference
  Series, Vol. 289, The Proceedings of the IAU 8th Asian-Pacific Regional
  Meeting, Volume 1, ed. S.~{Ikeuchi}, J.~{Hearnshaw}, \& T.~{Hanawa}, 291--294

\bibitem[{{Middleton} {et~al.}(2015){Middleton}, {Walton}, {Fabian}, {Roberts},
  {Heil}, {Pinto}, {Anderson}, \& {Sutton}}]{2015MNRAS.454.3134M}
{Middleton}, M.~J., {Walton}, D.~J., {Fabian}, A., {et~al.} 2015, \mnras, 454,
  3134, \dodoi{10.1093/mnras/stv2214}

\bibitem[{{Mineshige} {et~al.}(1994){Mineshige}, {Hirano}, {Kitamoto},
  {Yamada}, \& {Fukue}}]{1994ApJ...426..308M}
{Mineshige}, S., {Hirano}, A., {Kitamoto}, S., {Yamada}, T.~T., \& {Fukue}, J.
  1994, \apj, 426, 308, \dodoi{10.1086/174065}

\bibitem[{{Mitsuda} {et~al.}(1984){Mitsuda}, {Inoue}, {Koyama}, {Makishima},
  {Matsuoka}, {Ogawara}, {Shibazaki}, {Suzuki}, {Tanaka}, \&
  {Hirano}}]{1984PASJ...36..741M}
{Mitsuda}, K., {Inoue}, H., {Koyama}, K., {et~al.} 1984, \pasj, 36, 741

\bibitem[{{Mizuno} {et~al.}(2001){Mizuno}, {Kubota}, \&
  {Makishima}}]{2001ApJ...554.1282M}
{Mizuno}, T., {Kubota}, A., \& {Makishima}, K. 2001, \apj, 554, 1282,
  \dodoi{10.1086/321418}

\bibitem[{{Nasa High Energy Astrophysics Science Archive Research Center
  (Heasarc)}(2014)}]{2014ascl.soft08004N}
{Nasa High Energy Astrophysics Science Archive Research Center (Heasarc)}.
  2014, {HEAsoft: Unified Release of FTOOLS and XANADU}, Astrophysics Source
  Code Library, record ascl:1408.004.
\newblock \doeprint{1408.004}

\bibitem[{{Pakull} \& {Mirioni}(2002)}]{2002astro.ph..2488P}
{Pakull}, M.~W., \& {Mirioni}, L. 2002, arXiv e-prints, astro,
  \dodoi{10.48550/arXiv.astro-ph/0202488}

\bibitem[{{Pinto} {et~al.}(2016){Pinto}, {Middleton}, \&
  {Fabian}}]{2016Natur.533...64P}
{Pinto}, C., {Middleton}, M.~J., \& {Fabian}, A.~C. 2016, \nat, 533, 64,
  \dodoi{10.1038/nature17417}

\bibitem[{{Pinto} {et~al.}(2017){Pinto}, {Alston}, {Soria}, {Middleton},
  {Walton}, {Sutton}, {Fabian}, {Earnshaw}, {Urquhart}, {Kara}, \&
  {Roberts}}]{2017MNRAS.468.2865P}
{Pinto}, C., {Alston}, W., {Soria}, R., {et~al.} 2017, \mnras, 468, 2865,
  \dodoi{10.1093/mnras/stx641}

\bibitem[{{Pintore} {et~al.}(2017){Pintore}, {Zampieri}, {Stella}, {Wolter},
  {Mereghetti}, \& {Israel}}]{2017ApJ...836..113P}
{Pintore}, F., {Zampieri}, L., {Stella}, L., {et~al.} 2017, \apj, 836, 113,
  \dodoi{10.3847/1538-4357/836/1/113}

\bibitem[{{Poutanen} {et~al.}(2007){Poutanen}, {Lipunova}, {Fabrika},
  {Butkevich}, \& {Abolmasov}}]{2007MNRAS.377.1187P}
{Poutanen}, J., {Lipunova}, G., {Fabrika}, S., {Butkevich}, A.~G., \&
  {Abolmasov}, P. 2007, \mnras, 377, 1187,
  \dodoi{10.1111/j.1365-2966.2007.11668.x}

\bibitem[{{Prestwich} {et~al.}(2013){Prestwich}, {Tsantaki}, {Zezas},
  {Jackson}, {Roberts}, {Foltz}, {Linden}, \& {Kalogera}}]{2013ApJ...769...92P}
{Prestwich}, A.~H., {Tsantaki}, M., {Zezas}, A., {et~al.} 2013, \apj, 769, 92,
  \dodoi{10.1088/0004-637X/769/2/92}

\bibitem[{{Robba} {et~al.}(2021){Robba}, {Pinto}, {Walton}, {Soria}, {Kosec},
  {Pintore}, {Roberts}, {Alston}, {Middleton}, {Cusumano}, {Earnshaw},
  {F{\"u}rst}, {Sathyaprakash}, {Kyritsis}, \& {Fabian}}]{2021A&A...652A.118R}
{Robba}, A., {Pinto}, C., {Walton}, D.~J., {et~al.} 2021, \aap, 652, A118,
  \dodoi{10.1051/0004-6361/202140884}

\bibitem[{{Serlemitsos} {et~al.}(2007){Serlemitsos}, {Soong}, {Chan},
  {Okajima}, {Lehan}, {Maeda}, {Itoh}, {Mori}, {Iizuka}, {Itoh}, {Inoue},
  {Okada}, {Yokoyama}, {Itoh}, {Ebara}, {Nakamura}, {Suzuki}, {Ishida},
  {Hayakawa}, {Inoue}, {Okuma}, {Kubota}, {Suzuki}, {Osawa}, {Yamashita},
  {Kunieda}, {Tawara}, {Ogasaka}, {Furuzawa}, {Tamura}, {Shibata}, {Haba},
  {Naitou}, \& {Misaki}}]{2007PASJ...59S...9S}
{Serlemitsos}, P.~J., {Soong}, Y., {Chan}, K.-W., {et~al.} 2007, \pasj, 59, S9,
  \dodoi{10.1093/pasj/59.sp1.S9}

\bibitem[{{Shakura} \& {Sunyaev}(1973)}]{1973A&A....24..337S}
{Shakura}, N.~I., \& {Sunyaev}, R.~A. 1973, \aap, 24, 337

\bibitem[{{Shimura} \& {Takahara}(1995)}]{1995ApJ...445..780S}
{Shimura}, T., \& {Takahara}, F. 1995, \apj, 445, 780, \dodoi{10.1086/175740}

\bibitem[{{Soria} \& {Kong}(2016)}]{2016MNRAS.456.1837S}
{Soria}, R., \& {Kong}, A. 2016, \mnras, 456, 1837,
  \dodoi{10.1093/mnras/stv2671}

\bibitem[{{Thuan} {et~al.}(2004){Thuan}, {Bauer}, {Papaderos}, \&
  {Izotov}}]{2004ApJ...606..213T}
{Thuan}, T.~X., {Bauer}, F.~E., {Papaderos}, P., \& {Izotov}, Y.~I. 2004, \apj,
  606, 213, \dodoi{10.1086/382949}

\bibitem[{{Thuan} \& {Martin}(1981)}]{1981ApJ...247..823T}
{Thuan}, T.~X., \& {Martin}, G.~E. 1981, \apj, 247, 823, \dodoi{10.1086/159094}

\bibitem[{{Uchiyama} {et~al.}(2008){Uchiyama}, {Maeda}, {Ebara}, {Fujimoto},
  {Ishisaki}, {Ishida}, {Iizuka}, {Ushio}, {Inoue}, {Okada}, {Mori}, \&
  {Ozaki}}]{2008PASJ...60S..35U}
{Uchiyama}, Y., {Maeda}, Y., {Ebara}, M., {et~al.} 2008, \pasj, 60, S35,
  \dodoi{10.1093/pasj/60.sp1.S35}

\bibitem[{{Urquhart} \& {Soria}(2016)}]{2016MNRAS.456.1859U}
{Urquhart}, R., \& {Soria}, R. 2016, \mnras, 456, 1859,
  \dodoi{10.1093/mnras/stv2293}

\bibitem[{Virtanen {et~al.}(2020)Virtanen, Gommers, Oliphant, Haberland, Reddy,
  Cournapeau, Burovski, Peterson, Weckesser, Bright, {van der Walt}, Brett,
  Wilson, Millman, Mayorov, Nelson, Jones, Kern, Larson, Carey, Polat, Feng,
  Moore, {VanderPlas}, Laxalde, Perktold, Cimrman, Henriksen, Quintero, Harris,
  Archibald, Ribeiro, Pedregosa, {van Mulbregt}, \& {SciPy 1.0
  Contributors}}]{2020SciPy-NMeth}
Virtanen, P., Gommers, R., Oliphant, T.~E., {et~al.} 2020, Nature Methods, 17,
  261, \dodoi{10.1038/s41592-019-0686-2}

\bibitem[{{Walton} {et~al.}(2022){Walton}, {Mackenzie}, {Gully}, {Patel},
  {Roberts}, {Earnshaw}, \& {Mateos}}]{2022MNRAS.509.1587W}
{Walton}, D.~J., {Mackenzie}, A.~D.~A., {Gully}, H., {et~al.} 2022, \mnras,
  509, 1587, \dodoi{10.1093/mnras/stab3001}

\bibitem[{{Walton} {et~al.}(2018){Walton}, {F{\"u}rst}, {Heida}, {Harrison},
  {Barret}, {Stern}, {Bachetti}, {Brightman}, {Fabian}, \&
  {Middleton}}]{2018ApJ...856..128W}
{Walton}, D.~J., {F{\"u}rst}, F., {Heida}, M., {et~al.} 2018, \apj, 856, 128,
  \dodoi{10.3847/1538-4357/aab610}

\bibitem[{{Watarai} \& {Fukue}(1999)}]{1999PASJ...51..725W}
{Watarai}, K.-y., \& {Fukue}, J. 1999, \pasj, 51, 725,
  \dodoi{10.1093/pasj/51.5.725}

\bibitem[{{Watarai} {et~al.}(2000){Watarai}, {Fukue}, {Takeuchi}, \&
  {Mineshige}}]{2000PASJ...52..133W}
{Watarai}, K.-y., {Fukue}, J., {Takeuchi}, M., \& {Mineshige}, S. 2000, \pasj,
  52, 133, \dodoi{10.1093/pasj/52.1.133}

\bibitem[{{Watarai} \& {Mineshige}(2001)}]{2001PASJ...53..915W}
{Watarai}, K.-Y., \& {Mineshige}, S. 2001, \pasj, 53, 915,
  \dodoi{10.1093/pasj/53.5.915}

\bibitem[{{Wiktorowicz} {et~al.}(2017){Wiktorowicz}, {Sobolewska}, {Lasota}, \&
  {Belczynski}}]{2017ApJ...846...17W}
{Wiktorowicz}, G., {Sobolewska}, M., {Lasota}, J.-P., \& {Belczynski}, K. 2017,
  \apj, 846, 17, \dodoi{10.3847/1538-4357/aa821d}

\bibitem[{{Zampieri} \& {Roberts}(2009)}]{2009MNRAS.400..677Z}
{Zampieri}, L., \& {Roberts}, T.~P. 2009, \mnras, 400, 677,
  \dodoi{10.1111/j.1365-2966.2009.15509.x}

\bibitem[{{Zwicky}(1966)}]{1966ApJ...143..192Z}
{Zwicky}, F. 1966, \apj, 143, 192, \dodoi{10.1086/148490}

\end{thebibliography}
\bibliographystyle{aasjournal}

\end{document}